\documentclass[12pt]{article}
\usepackage[utf8]{inputenc}
\usepackage[margin=1.25in]{geometry}
\usepackage{natbib}
\usepackage{graphicx}
\usepackage{amsmath}
\usepackage{amsfonts}
\usepackage{setspace}
\usepackage{enumerate}   
\usepackage{enumitem}   
\usepackage{bbm}
\usepackage{adjustbox}
\usepackage{amsthm}
\usepackage{mathtools}
\usepackage[dvipsnames,table]{xcolor}
\usepackage{hyperref}
\usepackage{booktabs}
\usepackage{tabularx}
\usepackage{advdate}
\usepackage{environ}
\newcommand{\newinf}{\mathop{\mathrm{inf}\vphantom{\mathrm{sup}}}}
\def\D{\mathrm{d}}

\newlength{\myl}
\expandafter\let\expandafter\origequation\csname equation*\endcsname
\expandafter\let\expandafter\endorigequation\csname endequation*\endcsname
\long\def\[#1\]{\begin{equation*}#1\end{equation*}}
\RenewEnviron{equation*}{
  \settowidth{\myl}{$\displaystyle\BODY$} % calculate width and save as \myl
  \origequation
    \ifdim\myl>\linewidth
      \resizebox{\linewidth}{!}{$\displaystyle\BODY$}% \myl > \linewidth
    \else
      \BODY % \myl <= \linewidth
    \fi
  \endorigequation
}

\definecolor{darkgreen}{rgb}{0.0, 0.42, 0.24}

\newtheoremstyle{myplain}
  {\topsep}   % ABOVESPACE
  {\topsep}   % BELOWSPACE
  {\itshape\singlespacing}  % BODYFONT
  {0pt}       % INDENT (empty value is the same as 0pt)
  {\bfseries} % HEADFONT
  {.}         % HEADPUNCT
  {5pt plus 1pt minus 1pt} % HEADSPACE
  {}       
\theoremstyle{myplain}
\newtheorem{theorem}{Theorem}[section]

\newtheorem{proposition}{Proposition}[section]
\newtheorem{propcorollary}{Corollary}[proposition]
\newtheorem{lemma}{Lemma}[section]

\theoremstyle{definition}

\theoremstyle{definition}
\newtheorem{definition}{Definition}

\theoremstyle{remark}
\newtheorem{remark}{Remark}[section]

\numberwithin{equation}{section}

\title{\vspace{-1.7cm}Adversarial Estimators}
\author{Jonas Metzger \\ Stanford University}
\date{\AdvanceDate[-1]\today}

\begin{document}

\maketitle

\begin{abstract}
We develop an asymptotic theory of adversarial estimators (`A-estimators'). They generalize maximum-likelihood-type estimators (`M-estimators') as their average objective is maximized by some parameters and minimized by others. This class subsumes the continuous-updating Generalized Method of Moments, Generative Adversarial Networks and more recent proposals in machine learning and econometrics. In these examples, researchers state which aspects of the problem may {\it in principle} be used for estimation, and an adversary learns {\it how to emphasize} them optimally. We derive the convergence rates of A-estimators under pointwise and partial identification, and the normality of functionals of their parameters. Unknown functions may be approximated via sieves such as deep neural networks, for which we provide simplified low-level conditions. As a corollary, we obtain the normality of neural-net M-estimators, overcoming technical issues previously identified by the literature. Our theory yields novel results about a variety of A-estimators, providing intuition and formal justification for their success in recent applications.
\end{abstract}

\onehalfspacing
\section{Introduction}

Although it is not always obvious, nearly all population parameters that are estimated in econometrics and machine learning can be written as the solution of so-called saddle-point or adversarial objectives of the form:
\begin{equation}
\label{def:Aestimand}
    \theta_* = \underset{\theta \in \Theta}{\arg\min} \max_{\lambda\in\Lambda} \mathbb{E} l(\theta, \lambda, Y)
\end{equation}
where $l$ is a known loss function, $Y$ is a random variable and $\Theta, \Lambda$ are parameter spaces, containing the unknown parameter of interest $\theta_*$ and nuisance $\lambda$. We examine the natural estimator $\widehat\theta_n$ that approximately solves the empirical Nash condition:
\begin{align}
    \label{assn:outernash}\mathbb{E}_n l(\widehat\theta_n, \widehat\lambda_n, Y) & \leq \inf_{\theta\in\Theta_n} \mathbb{E}_n l(\theta, \widehat\lambda_n, Y) + \widetilde\eta_n  \\
    \label{assn:innernash}\mathbb{E}_n l(\widehat\theta_n, \widehat\lambda_n, Y) & \geq \sup_{\lambda\in\Lambda_n} \mathbb{E}_n l(\widehat\theta_n, \lambda, Y) - \eta_n
\end{align}
which replaces the expectation $\mathbb{E}$ of the population objective \ref{def:Aestimand} with the average of $n$ iid samples, $\mathbb{E}_n$. We search for the estimators over so-called {\it sieve spaces}  $\widehat\theta_n,\widehat\lambda_n\in\Theta_n,\Lambda_n$ (\cite{grenander1981abstract}), which approximate the full parameter spaces $\Theta_n,\Lambda_n\subset\Theta,\Lambda$ and grow with the sample size $n$. These could be neural networks for example, growing in depth and width. The sequences $\widetilde \eta_n,\ \eta_n=o_\mathbb{P}(1)$ accommodate numerical procedures which only yield approximate Nash equilibria. This class of A-estimators (A for {\it adversarial}) strictly generalizes so-called M-estimators (M for {\it maximum likelihood-type}), which are obtained by fixing $\Lambda$ to be singleton.\\

A-Estimators have become a workhorse of econometrics and causal inference long before the advent of deep learning. \cite{CUEGMM}'s continuous-updating Generalized Methods of Moments (GMM), which looks for $\theta$ satisfying $\mathbb{E}[m(\theta,Y)]=0$ for some known function $m(\theta, Y)$, can be written as:
\begin{align*}
\inf_\theta  \mathbb{E}_n\left[m(\theta,Y)\right] &\mathbb{E}_n\left[m(\theta,Y)m(\theta,Y)'\right]^{-1}\mathbb{E}_n\left[m(\theta,Y)\right] \\ &=\inf_\theta\sup_\lambda\mathbb{E}_n\left[m(\theta,Y)'\lambda - (m(\theta,Y)'\lambda)^2/4\right]
\end{align*}
and is therefore an A-estimator, but not an M-estimator. In statistics, an earlier example consists of the Empirical Likelihood (EL) approach pioneered in \cite{cosslet1981, owen1988empirical, owen1990empirical, qin1994empirical}. Subsequently, EL was unified with GMM into the Generalized Empirical Likelihood (GEL) framework (\cite{newey2004generalized}), also subsuming the exponential-tilting estimator (\cite{imbens1995gel}), for example. All GEL estimators are A-estimators, but their adversarial formulation was rarely salient. However, some of their benefits may be owed directly to their adversarial objective: the adversary $\lambda$ automatically detects which moment violations are most informative at a given parameter guess, adaptively guiding the estimation towards an efficient solution. This contrasts with earlier estimators which weighted the moments in a way that depended on choices of the researcher: the weights of Pearson's Method-of-Moments were manually set by the researcher (implicitly), resulting in inefficient root-$n$ asymptotics. Two-step GMM (\cite{hansen1982gmm}) required choosing a first-step estimator to compute the weights, yielding inefficient higher-order asymptotics (see \cite{newey2004generalized}). Formally, the optimal weights are nuisance parameters, and as we will see in Section \ref{subsec:neyman}, estimating them via an adversary ensures that $\widehat{\theta}_n$ is robust to estimation errors in these nuisance parameters.\\

A key invention which put a spotlight on adversarial objectives in recent years were \textit{Generative Adversarial Networks}, or GANs (\cite{goodfellow2014generative}). They search for a generative model $Y\sim\mathbb{P}_\theta$ for which no adversary $\lambda(Y)\in(0,1)$ (called `critic' or `discriminator') could tell apart the generated data from $n$ real samples:
$$\inf_{\mathbb{P}_\theta}\sup_{\lambda(\cdot)\in(0,1)} \mathbb{E}_{\mathbb{P}_\theta} \log\lambda(Y)+\mathbb{E}_n\log(1-\lambda(Y))$$
The objective contains the log-likelihood of a binary classifier $\lambda(Y)$ discriminating between an equal number of real and generated samples. As we show in Section \ref{ex:fDiv}, this directly measures the Jensen-Shannon divergence between $\mathbb{P}_\theta$ and $\mathbb{P}_n$. As of today, 
versions of this objective are key to state-of-the-art image generation, see e.g. \cite{Jabbar2022ASO} for a recent survey. An analogy to human-generated images makes this unsurprising: it is much easier to tell apart a photo from an image drawn by a human, than it is to draw a realistic image, or to define what makes a drawing realistic. This intuition motivates the objective: train the generator until its critic has nothing more to criticize. The ingenuity is that the researcher need not define a meaningful measure of `realism' of a piece of data anymore. Instead, this measure is \textit{learned} by the adversary. It is clear that the utility of this idea extends beyond image generation: in Imitation Learning, a sub-field of Robotics, it has been used to teach human behavior to artificial agents without requiring hand-crafted measures of `humanness' (\cite{gail}). In Econometrics, where new causal inference methods can usually only be benchmarked on simulated data sets, \cite{athey2019using} used the objective to limit the impact of researcher's subjective choices by requiring simulations to be indistinguishable from real data. \cite{kaji2020adversarial} proposed to use the objective to estimate structural economic models which produce realistic data beyond the set of features that would otherwise be manually specified by the researcher.\\

More generally, other adversarial objectives have proven useful beyond fitting models to data. In Reinforcement Learning, a sub-field of Robotics where agents independently discover strategies to reach predefined goals without copying prior examples, \cite{sbeed} proposed an A-estimator in which the adversary detects and penalizes any systematic deviation from optimal behavior. \cite{constrainedOpt} proposed an estimator which extends a standard ML objective by an adversary imposing fairness constraints across sub-populations. More recently, research in econometrics established A-estimators as a natural framework for integrating machine learning methods into causal inference, where quantities of interest are frequently identified by a continuum of restrictions. \cite{chernozhukov2020adversarial} propose to estimate Riesz representers of causal parameters directly, via an adversary enforcing the restrictions identifying the Riesz representer. Estimating Riesz representers is key to obtaining well-behaved estimates of causal parameters in the presence of nuisance functions, and can also be useful for estimating asymptotic variances, e.g. \cite{penalizedsieve2019}. Another line of research develops novel adversarial objectives to estimate causal parameters from {\it conditional} moment restrictions, which naturally arise from causal assumptions (e.g. the instrumental variable setting), and are usually more informative than any finite set of unconditional moment restrictions. In this line of research, the adversary can be viewed as adaptively finding the unconditional moment restriction which is most violated at the current parameter guess, among infinitely many which are implied by the conditional moment restriction. The key works are \cite{adversarialGMM2018,dikkala2020minimax} and \cite{deepGMM2019, variationalMM2020}. \cite{metzger2021} propose a semi-parametrically efficient generalization of GEL to the conditional case via adversarial networks, containing \cite{variationalMM2020} as a special case. \\

In summary, a recurring theme of adversarial objectives is that instead of manually defining which specific features of the data are important for a model to capture, the researcher's role is restricted to stating a general principle which should be satisfied by all features of the correct model, and the adversary {\it adaptively} focuses the estimation on the model's features which violate this principle the most. Over the course of the paper, we will encounter further interesting connections between various A-estimators, such as their Neyman orthogonality, their information-theoretic foundation via f-Divergences, and their ties to Lagrangian Duality.\\

Despite their popularity, we are not aware of a unified statistical theory of A-estimators. For some individual estimators, consistency (\cite{deepGMM2019}) and convergence rate results (\cite{dikkala2020minimax, singh2018nonparametric,liang2018HowWell,belomestny2021rates}) were obtained, but normality results are limited to parametric $\theta$, either in Kernel settings (\cite{variationalMM2020}) or leaving high-level assumptions about neural networks unverified (\cite{kaji2020adversarial}). This can be attributed to two main obstacles: the theory of M-estimation does not apply to A-estimators, and the arguments from which the former is built up are insufficient to e.g. obtain the required uniform convergence of the adversary. The second issue is that adversarial objectives are most popular in the context of (deep) neural networks, whose statistical analysis (particularly their asymptotic normality) is complicated, e.g. due to their non-convex sieve space. Even in M-estimation settings, it was not clear whether known, high-level conditions for normality could be verified for neural networks (cf. the Conclusion of \cite{shen2019asymptotic}). We therefore make three separate contributions:
\begin{enumerate}
    \item We characterize the general class of A-estimators, and show that a wide range of estimators proposed in econometrics and machine learning fall into this class. We point out desirable characteristics shared between A-estimators, which help explain their recent success in practice.  
    \item We develop a unified statistical theory of A-estimators, yielding their consistency, convergence rates (both under point- and partial identification), and asymptotic normality of functionals of their parameters. We provide high-level conditions for arbitrary sieves, as well as low-level conditions for semi-parametric settings with neural networks, to simplify verification in practice.
    \item  We extend the theory of neural network M-estimators (as a special case). Our convergence rates hold uniformly over families of losses, allow more general losses than \cite{farrell2018deep} and attain a reduced curse-of-dimensionality which \cite{minkowski, bauer2019deep} observed in regression settings with lower-dimensional structures. To the best of our knowledge, we provide the first normality result for functionals of deep neural networks which does not rely on Neyman-orthogonality or unverified high-level assumptions.
\end{enumerate}

The remainder of the paper is structured as follows. In Section \ref{sec:examples}, we review five different A-estimators proposed in the econometrics and machine learning literatures. We present our general statistical theory of A-estimators in Section \ref{sec:theory} and apply it in Section \ref{sec:application} to derive novel results about the examples of Section \ref{sec:examples}. We conclude by recapping the similar role adversaries play across all examples, providing intuition which types of problems may generally benefit from adversarial formulations. Appendix \ref{appendix:proofs} and Online Appendix \ref{appendix:online} contain the proofs omitted in Sections \ref{sec:theory} and \ref{sec:application}, respectively.\looseness=-1

%On the machine learning side, for the broad class of GANs minimizing an f-divergence (\cite{nowozin2016fgan}, subsuming the original \cite{goodfellow2014generative}), we derive convergence rates exhibiting a reduced curse-of-dimensionality, the asymptotic normality of smooth functionals of their parameters, and provide conditions for their efficiency with semiparametric $\theta$ - results which hitherto did not exists in the literature at this level of generality. For the popular off-policy reinforcement learning algorithm of \cite{sbeed}, we derive significantly faster convergence rates than the original the original work. We then examine three econometric estimators: for Generalized Empirical Likelihood (specifically cuGMM), we use our theory to derive the convergence rates in a novel partially identified setting where neural nets approximate unknown functions. Next, we show that the neural-network based conditional moment estimator of \cite{dikkala2020minimax} attains a reduced curse of dimensionality as well. Unlike the original work, we further derive its asymptotic distribution, showing that it is inefficient, and discuss how to modify it to attain efficiency. Finally, we complement \cite{chernozhukov2020adversarial}'s proposed Riesz representer A-estimator, by showing that using neural networks can weaken the smoothness and dimensionality restrictions required for downstream causal inference.
\section{Examples}\label{sec:examples}

\subsection{Minimum $f$-Divergence}
\label{ex:fDiv}

A powerful class of estimation objectives asymptotically minimize an $f$-divergence $D_f(\mathbb{P}_\theta \| \mathbb{P})$ between the distribution of the data $Y\sim \mathbb{P}=\mathbb{P}_{\theta_*}$ and the distribution of some model $\mathbb{P}_\theta, \theta\in\Theta_n$ with support $\mathcal{Y}$. This class, introduced by \cite{nowozin2016fgan}, subsumes GANs (\cite{goodfellow2014generative}), and many follow ups such as \cite{mao2017squares}, \cite{chisqgan}. For a continuous, proper convex function $f: \mathbb{R}\mapsto \mathbb{R}$ satisfying $f(1)=0$, the $f$-divergence is defined as $D_f(\mathbb{P}_\theta \| \mathbb{P})=\mathbb{E}_\mathbb{P}[f(\frac{\D\mathbb{P}_\theta(Y)}{\D\mathbb{P}(Y)})]$, where $\frac{\D\mathbb{P}_\theta(Y)}{\D\mathbb{P}(Y)}$ denotes the Radon-Nikodym derivative of $\mathbb{P}_\theta$ with respect to $\mathbb{P}$ (=likelihood ratio), which we assume exists for all $\theta\in\Theta$. Notably, $D_f(\mathbb{P}_\theta \| \mathbb{P})$ admits a useful dual representation:
\begin{equation}
\label{eq:fdiv}
    D_f(\mathbb{P}_\theta \| \mathbb{P}) := \mathbb{E}_\mathbb{P} f\left(\frac{\D\mathbb{P}_\theta(Y)}{\D\mathbb{P}(Y)}\right) = \sup_{\lambda: \mathcal{Y} \to \mathbb{R}} \mathbb{E}_{\mathbb{P}_\theta}[\lambda(Y)] -  \mathbb{E}_\mathbb{P}[f_*(\lambda(Y))]
\end{equation}
where $f_*(t) := \sup_{\lambda\in\mathbb{R}} \lambda t-f(\lambda)$ denotes the {\it convex conjugate} of $f$. The equality above follows from $f=(f_*)_*$. Various choices\footnote{None of the objectives are unique: $f(t) \gets f(t)+c(t-1)$ for any $c$ yields the same divergence, but changes the expressions. Note that we may also swap $\mathbb{E}_{\mathbb{P}_\theta}$ and $\mathbb{E}_n$, which yields valid objectives for the respective ``reverse'' $f$-divergences.} for $f$ are presented in Table \ref{tab:fdiv}. This duality is useful because the right-hand side suggests a finite-sample analog which does not depend on unknown quantities: we obtain an A-estimator for $\theta_*$ by letting
\begin{equation}
\label{eq:fdivloss}
    l(\theta, \lambda, Y) = \mathbb{E}_{\mathbb{P}_{\theta}}[\lambda(Y)] -  f_*(\lambda(Y))
\end{equation}
and solving for $\widehat\theta_n, \widehat\lambda_n$ satisfying the Nash condition \ref{assn:outernash},\ref{assn:innernash} in $\mathbb{E}_n[l(\theta, \lambda, Y)]$. Normalizing $f(t) \gets \frac{f(t)-f'(1)(t-1)}{f''(1)}$ without loss of generality\footnote{This implies $f'(1)=0$, $f''(1)=1$ and $f_*(0)=0, f_*'(0)=f_*''(0)=1$, which merely re-scales the divergence \ref{eq:fdiv} by a factor of $1/f''(1)$}, assuming the second derivative $f''$ exists, the function $\lambda$ attaining the supremum in \ref{eq:fdiv} at some $\theta$ is $\lambda_*^\theta=f'(\frac{\D\mathbb{P}_\theta}{\D\mathbb{P}})$. The adversary $\widehat\lambda_n$ therefore estimates this transformed likelihood ratio at the current guess for $\widehat\theta_n$, and the Nash-equilibrium corresponds to the case where it is approximately constant, i.e. the distribution $\mathbb{P}_{\widehat\theta_n}$ is close to that of the data. Notably, $\mathbb{E}_n[l(\theta, \lambda, Y)]$ can be evaluated using only samples from the two distributions\footnote{Note that we neither require explicit knowledge of $\mathbb{P}_{\theta}$ nor infinitely many samples from $\mathbb{P}_{\theta}$ at a given $n$: it suffices to draw $m \succ n^2$ Monte Carlo samples from $\mathbb{P}_{\theta}$ and solve for the corresponding finite sample saddle point. The resulting Monte Carlo approximation error for the expectation $\mathbb{E}_{\mathbb{P}_{\theta}}$ is then of order $\sqrt{m}^{-1}=n^{-1}$ and can thus be accounted for by letting $\widetilde\eta_n, \eta_n=O_\mathbb{P}(n^{-1})$ in equations \ref{assn:outernash},\ref{assn:innernash}, which has no impact on our asymptotic results.}. This is crucial for GANs, where $\mathbb{P}_\theta$ is only implicitly defined via a push-forward mapping parametrized by a neural net. As proposed by \cite{kaji2020adversarial}, this also makes it a drop-in alternative to the {\it Simulated Method of Moments}, which similarly estimates economic models from data they generate, but matches only a finite set of moments instead of the full distribution. \\

\begin{table}[h!]
\label{tab:fdiv}
\begin{adjustbox}{max width=\textwidth}
\begin{tabular}{l|l|l|l}
Name                & $f(t)$    & $f_*(t)$ , domain            & Generative Adversarial Objective for $\theta$  \\ \hline
Total Variation     & $|t-1|/2$ & $t$, for $|t| \leq \frac{1}{2}$  & $\sup_{|\lambda|\leq \frac{1}{2}} \mathbb{E}_{\mathbb{P}_\theta} \lambda(Y)-\mathbb{E}_n \lambda(Y)$ \\
KL Divergence       & $t\log t$ & $e^{t-1}$                                    & $\sup_{\lambda\in\mathbb{R}} 1 + \mathbb{E}_{\mathbb{P}_\theta} \lambda(Y)-\mathbb{E}_n e^{\lambda(Y)}$                                                                                                                      \\
Reverse KL       & $-\log t$ & $-\log(-te)$, for $t\leq0$       & $\sup_{\lambda\leq 0} 1 + \mathbb{E}_{\mathbb{P}_\theta} \lambda(Y)+\mathbb{E}_n \log(-\lambda(Y))$    \\
$\chi^2$ Divergence & $(t-1)^2$ & $t+t^2/4$                                    & $\sup_{\lambda\in\mathbb{R}} \mathbb{E}_{\mathbb{P}_\theta} \lambda(Y)-\mathbb{E}_n \left[\lambda(Y) + \lambda(Y)^2/4 \right]$        \\
Squared Hellinger & $(\sqrt{t}-1)^2$ & $\frac{t}{1-t}$, for $ t\leq 1$ & $\sup_{\lambda\leq 1} \mathbb{E}_{\mathbb{P}_\theta} \lambda(Y)-\mathbb{E}_n \left[\frac{\lambda(Y)}{1-\lambda(Y)}\right]$ \\
rescaled JS (GAN) & $t\log t-(1+t)\log(1+t)$ &$\begin{matrix}\small{-\log(1-e^t)}, \text{for } t<0\end{matrix}$ &  $\sup_{\log\lambda<0} \mathbb{E}_{\mathbb{P}_\theta} \log\lambda(Y)+\mathbb{E}_n\log(1-\lambda(Y))$ 
\end{tabular}
\end{adjustbox}
\caption{Various adversarial $f$-divergence objectives. $f_*(t)=\infty$ outside the domain.}
\end{table}

\subsection{Generalized Empirical Likelihood} \label{ex:GEL}
Our next example is a class of A-estimators that was proposed long before the recent success of adversarial objectives in deep learning. In econometrics, many important parameters $\theta_*$ are identified by a moment restriction of the form:
$$\mathbb{E}[m(Y, \theta)]=0 \ \iff \ \theta=\theta_*$$
for some known, possibly vector-valued function $m(Y,\theta)$. In the Introduction, we presented the continuous-updating GMM objective (\cite{CUEGMM}) for estimating $\theta_*$, a workhorse for causal inference in econometrics. In this section, we review the more general class of Generalized Empirical Likelihood (GEL) estimators (\cite{newey2004generalized}), which solve the constrained minimization problem:
$$\inf_{\Bar{\mathbb{P}},\theta\in\Theta} D_f(\Bar{\mathbb{P}}\|\mathbb{P}_n)  \text{ s.t. } \mathbb{E}_{\Bar{\mathbb{P}}}[m(Y,\theta)]=0$$
That is, they seek for a parameter $\theta$ and a corresponding population distribution $\Bar{\mathbb{P}}$ that is as close as possible to the sample $\mathbb{P}_n$, subject to satisfying the moment constraint $\mathbb{E}_{\Bar{\mathbb{P}}}[m(Y,\theta)]=0$. At this high level, it is worth noting that GEL optimizes the same target as the objective in Section \ref{ex:fDiv}, which imposes $\Bar{\mathbb{P}}=\mathbb{P}_\theta$ instead of a moment constraint. Glossing over some details, we can obtain a tractable estimator in this setting by concentrating out $\Bar{\mathbb{P}}$ from the corresponding Lagrangian:
\begin{align}
    \begin{split}\label{eq:gel}
        \inf_{\Bar{\mathbb{P}},\theta\in\Theta} \sup_{\lambda\in\mathbb{R}^{\operatorname{dim}(m)}} & D_f(\Bar{\mathbb{P}}\|\mathbb{P}_n) - \lambda'\mathbb{E}_{\Bar{\mathbb{P}}}[m(Y,\theta)] = \inf_{\theta\in\Theta} \sup_{\lambda\in\mathbb{R}^{\operatorname{dim}(m)}} \mathbb{E}_n\left[-f_*\left(\lambda'm(Y, \theta)\right)\right]
    \end{split}
\end{align}
Which again uses the convex conjugate $f_*$ of $f$ (see example \ref{ex:fDiv}). For a formal proof of this equivalence, see e.g. \cite{imbens1995gel}. It is easy to see that GEL is an A-estimator with $l(\theta, \lambda , Y)=-f_*(\lambda'm(Y, \theta))$ where $\Lambda_n=\mathbb{R}^{\operatorname{dim}(m)}$ and $\Theta_n$ is the parameter space of the economic model. A particularly popular version of this objective corresponds to the case $D_f=\chi^2$, where Table \ref{tab:fdiv} tells us that $f(t)=(t-1)^2$ and $f_*(t)=t+t^2/4$. In this case, we can analytically solve for the optimal adversary given $\theta$. Substituting it in, we get the continuous-updating GMM objective presented in the introduction:
$$\sup_{\lambda\in\mathbb{R}^{\operatorname{dim}(m)}} \mathbb{E}_n\left[-f_*\left(\lambda'm(Y, \theta)\right)\right] = \mathbb{E}_n\left[m(Y, \theta)\right]'\mathbb{E}_n\left[m(Y, \theta)m(Y, \theta)'\right]^{-1}\mathbb{E}_n\left[m(Y, \theta)\right]$$

\subsection{Off-Policy Reinforcement Learning} \label{ex:RL}

Next, we review the Smoothed Bellman Error Embedding (SBEED) algorithm introduced by \cite{sbeed}, a popular off-policy learning method in robotics. Off-policy learning aims to learn the optimal policy for an agent from data that was generated under an entirely different policy regime. This problem is not limited to robotics: since it was identified in the monetary policy context by \cite{LUCAS197619}, it became a primary concern in econometrics and its recognition played a key role in the {\it credibility revolution} (\cite{credibilityrevolution}) of econometrics. While problem definitions otherwise differ between these literatures, off-policy learning methods have received recent interest in econometrics (\cite{offpolicyathey, policylearning}).\\

For an agent receiving reward $R(s,a)$ for taking action $a\in\mathcal{A}$ at state $s\in\mathcal{S}$, forming an expectation over the future state $s^{+}\in\mathcal{S}$, SBEED's goal is to learn the value function $V_*(s)$ and policy $a\sim P_*(\cdot|s)$  which satisfy the regularized Bellman equation:
$$V_*(s)=\max_{P(\cdot|s)} \mathbb{E}_{a\sim P(\cdot|s)}\Big[R(s,a)+\beta \mathbb{E}_{s^+|s, a}[V_*(s^+)|s, a]\Big] + H(P, s)$$
where the entropy $H(P, s)=-\mathbb{E}_{a\sim P(\cdot|s)}[\log P(a|s)]$ regularizes the optimal policy $P_*(\cdot|s)$ towards exploring all actions $a\in\mathcal{A}$. Given the researcher's choice of $R, \beta$, the goal is to learn $P_*, V_*$ from finite samples $\{(s_i, a_i, s^+_i)\}_{i=1}^n$. Importantly, the actions $a_i$ may be sampled from a {\it suboptimal} policy which does not equal $P_*$. Starting from the first-order condition of the Bellman equation, \cite{sbeed} develop an adversarial population objective, whose finite-sample analog is the A-estimator \ref{assn:outernash},\ref{assn:innernash} with loss:
\begin{equation}\label{eq:sbeed}
  l(\theta, \lambda, Y) = \big(R(s,a) +\beta V_\theta(s^+) - V_\theta(s)-\log P_\theta(a|s)\big)\lambda(s, a) - \frac{1}{2}\lambda(s, a)^2  
\end{equation}
where $\lambda(s, a)$, $\log P_\theta(a|s)$ and $V_\theta(s)$ are implemented as neural networks in practice. 

\subsection{A-Estimators for Conditional Moment Restrictions} \label{ex:CMR}

Another powerful application for A-estimators recently pursued by the econometric literature are conditional moment estimators. These methods estimate parameters $\theta_*$ which are identified by restrictions of the form:
\begin{equation}\label{eq:cmr}
    \mathbb{E}[m(X, \theta)|Z]=0 \ \forall Z \iff \ \theta=\theta_*
\end{equation}
for some random variables $Y=(X,Z)$ and a known function $m(X,\theta)$. Conditions of this type occur e.g. when estimating some causal effect $\theta$ via instrumental variables, or as the first-order conditions of agents optimizing some expected utility given some information $Z$. As a result, nonparametric conditional moment estimators received considerable interest in econometrics, see e.g. \cite{aichen2003, aichen2007, chenqiu2016}. These earlier estimators rely on first-step estimates of nuisance parameters capturing the conditional means and variances. Intuitively however, estimating the nuisance parameters via predictive objectives in a separate first step may dedicate scarce model capacity to capturing features which are not useful for the purpose of estimating $\theta_*$ downstream. This motivates recent work on adversarial objectives which unify the estimation into a single objective, more plausibly targeting the nuisance estimation towards the goal of identifying $\theta_*$. Specifically, we will examine the estimator of \cite{dikkala2020minimax}, with $l(\theta,\lambda,Y)=m(X, \theta)'\lambda(Z)-\frac{1}{4}\|\lambda(Z)\|_2^2$, yielding the finite sample objective
$$\inf_{\theta\in\Theta_n}\sup_{\lambda\in\Lambda_n}  \mathbb{E}_n\left[ m(X,\theta)'\lambda(Z) -  \frac{1}{4} \lambda(Z)'\lambda(Z) \right],$$
where $\Lambda_n$ is a class of neural networks. The methods proposed by \cite{Bennett2019DeepGM, variationalMM2020} are closely related, but differ in the penalty they impose on $\lambda$. \cite{dikkala2020minimax} consider the case of instrumental variable regression, where $X=(y,x)$ and $m(X,\theta)=y-\theta(x)$, but we will examine the general case. We note that Example \ref{ex:RL} (SBEED, \cite{sbeed}) can be viewed as a special case of re-scaled version of this objective, with $X=(s,a, s^+)$ and $Z=(s,a)$, although both literatures seem to be unaware of their connection. One can analytically solve for the optimal adversary $\lambda_*^\theta(Z)=2\mathbb{E}[m(X,\theta)|Z]$ to rewrite the population objective as:
$$\mathbb{E}[l(\theta,Y)]:=\mathbb{E}[l(\theta, \lambda_*^\theta,Y)] =\mathbb{E}[\|\mathbb{E}[m(X,\theta)|Z]\|_2^2]=\mathbb{E}[\|\mathbb{E}[m(X,\theta)-m(X,\theta_*)|Z]\|_2^2]$$
which can be understood as a measure of distance between $\theta$ and $\theta_*$, which clearly attains its minimum at $\theta=\theta_*$, when $\mathbb{E}[m(X,\theta)|Z]\equiv0$. In Section \ref{app:CMR}, we will apply our theory to derive the asymptotic distribution of this estimator and show that is in fact {\it inefficient}. We further discuss how the adversarial formulation of GMM can directly inform a simple modification similar to \cite{variationalMM2020} which yields an efficient A-estimator.

\subsection{Estimating Riesz Representers}
\label{ex:riesz}

\cite{chernozhukov2020adversarial} propose a distinct A-estimator to estimate {\it Riesz representers} for structural parameters $\phi_*$ which can be written as linear functionals $\phi_*=\phi(g_*)=\mathbb{E}[m(Y,g_*)]$. Here, $g_*=\mathbb{E}[y|x]$ is an unknown function for which an estimate $\widehat{g}_n$ is available from some first-stage regression of $y$ on $x$, where $Y=(y,x,w)$. Quantities like $\phi_*$ are common in the average treatment effect or asset pricing literature, for example. Unfortunately, especially if $\widehat{g}_n$ is estimated via machine learning, the `naive' estimator 
$$\widehat{\phi}_n=\mathbb{E}_n[m(Y,\widehat{g}_n)]$$
is often not well behaved: $\sqrt{n}(\widehat{\phi}_n-\phi_*)$ may not converge in distribution to a Gaussian limit and thus one cannot provide confidence intervals around the estimate. Under the conditions of the Riesz representation theorem however, there may exist a function $\theta_*\in\Theta$ called the {\it Riesz representer} of the functional $\phi(g)$, which satisfies:
$$\phi(g)=\mathbb{E}[\theta_*(x)g(x)] \quad \forall g\in\Theta$$
If a well-behaved estimate $\widehat\theta_n$ of $\theta_*$ is available, it can be combined with $\widehat{g}_n$ to define the so-called {\it orthogonalized} estimator:
$$\Tilde\phi_n = \mathbb{E}_n[m(Y,\widehat{g}_n)-\widehat\theta_n(x)(y-\widehat{g}_n(x))]$$
which attains asymptotic normality under rather weak conditions on $\widehat{g}_n$ (see Lemma 17 of \cite{chernozhukov2020adversarial}). \cite{chernozhukov2020adversarial} propose a generalized procedure to estimate $\widehat\theta_n$ via an A-estimator, which we will simplify as follows:
$$\inf_{\theta\in\Theta_n}\sup_{\lambda\in\Lambda_n}\mathbb{E}_n[m(Y,\lambda)-\theta(x)\lambda(x)-\lambda(x)^2/2]$$
where $\Theta_n,\Lambda_n$ are neural networks. To clarify why this objective works, is it useful to analytically solve the adversarial component of the corresponding population objective: $$\sup_\lambda\mathbb{E}[m(Y,\lambda)-\theta(x)\lambda(x)-\lambda(x)^2/2]=\frac{1}{2}\mathbb{E}[(\theta_*(x)-\theta(x))^2]$$
As we will show in Section \ref{app:riesz}, our theory directly yields the convergence rates for $\widehat\theta_n$ that \cite{chernozhukov2020adversarial}'s Lemma 17 requires for the asymptotic normality of $\Tilde\phi_n$. It does so at a reduced curse of dimensionality in $x$ for rather general function classes - i.e. under weaker conditions on smoothness and dimension of the data - complementing the original work.

\section{General Theory}
\label{sec:theory}
\paragraph{Roadmap.} This Section will present our general theory of A-estimators. Subsection \ref{subsec:nashvsminimax} briefly discusses an alternative definition of A-estimators that may be more natural to some readers. In Subsection \ref{subsec:neyman}, we establish that A-estimators satisfy the desirable condition of Neyman-orthogonality with respect to the adversary and discuss its implications. Next, we characterize the convergence rates of A-estimators: Section \ref{subsec:generalrate} provides a high-level result for arbitrary sieves such as splines or wavelets, not just neural nets. Under more easily verifiable low-level conditions, Subsection \ref{subsec:neuralArate} provides convergence rates for semiparametric settings involving neural networks, showing they exhibit a reduced curse-of-dimensionality. Finally, we characterize the asymptotic normality of smooth functionals of A-estimators. We again begin with a general, high-level result for arbitrary sieves, followed with the low-level conditions for the normality of neural networks. Notably, we show that a combination of undersmoothing and regularizing towards a convex target space suffices to overcome a key issue for normality proofs of neural networks: their non-convex sieve space.

\paragraph{Notation.}
Throughout, we consider random variable $Y$ with support $\mathcal{Y}$, distribution $\mathbb{P}$ and corresponding expectation operator $\mathbb{E}$. We also denote the variance operator by $\mathbb{V}[f(Y)]=\mathbb{E}(f(Y)-\mathbb{E}[f(Y)])^2$ for any function $f:\mathcal{Y}\mapsto\mathbb{R}$. We denote the sample average, i.e. the expectation under the empirical distribution $\mathbb{P}_n$, by $\mathbb{E}_n$. Throughout, $\mathbb{E}$ will treat estimated parameters as deterministic sequences indexed by $n$, as is common in the literature. We also consider subvectors of $Y$, denoted by $x\in {\mathcal{X}}, \Bar x\in \Bar {\mathcal{X}}$, with their respective supports ${\mathcal{X}}, \Bar {\mathcal{X}}$ being subspaces of $\mathcal{Y}$. We require various norms: throughout, $\|x\|_q$ will denote the $\ell^q$ norm of a finite dimensional vector $x$, with $\|x\|=\|x\|_2$ being the Euclidean norm. For a possibly vector-valued function $f(x)$, we denote its $L^q$ function norm over some subset $\widetilde{\mathcal{X}}\subset\mathcal{X}$ by $\|f\|_{L^q(\widetilde{\mathcal{X}})}=\mathbb{E}[\|f(x)\|^q_q|x\in \widetilde{\mathcal{X}}]^{1/q}$. We denote the supremum norm of a vector $x$ with components $x_i$ by $\|x\|_\infty=\max_i |x_i|$. The supremum norm of $f$ over $\widetilde{\mathcal{X}}$ will be denoted by $\|f\|_{\widetilde{\mathcal{X}}}=\sup_{x\in {\widetilde{\mathcal{X}}}} \|f(x)\|_\infty$. For $\widetilde{\mathcal{X}}=\mathcal{X}$, we may omit the dependence on ${\mathcal{X}}$ by writing $\|f\|_\infty:=\|f\|_{\mathcal{X}}$. We will often write $a\prec b$ to denote $a=O(b)$, implying that a sufficiently large global constant $\infty>C>0$ exists such that $a\leq C b$, where $C$ does not depend on any varying aspects of the problem, such as any parameters, sample sizes, et cetera. We write $a\asymp b$ if $a\prec b \prec a$. We will also write $a\vee b = \max(a, b)$ and $a \wedge b = \min(a,b)$. Throughout, we will write $l^\theta(\lambda, Y)=l(\theta, \lambda, Y)$ and $l(\theta, Y)=l(\theta, \lambda_*^\theta, Y)$ for short, where $\lambda_*^\theta=\arg\max_{\lambda\in\Lambda} \mathbb{E} l(\theta, \lambda, Y)$. We denote by $\pi_n$ a (not necessarily linear) projection onto the respective sieves, i.e. $\pi_n \theta\in\arg\inf_{\theta'\in\Theta_n} \|\theta'-\theta\|_\infty$ for any $\theta\in\Theta$ and $\pi_n \lambda\in\arg\inf_{\lambda'\in\Lambda_n} \|\lambda'-\lambda\|_\infty$ for any $\lambda\in\Lambda$.

\subsection{Nash vs Minimax}\label{subsec:nashvsminimax}

We presented our preferred definition for A-estimators in the introduction, as satisfying a Nash condition of the empirical loss. All results of this paper will apply to this definition. However, the reader may have noticed that the ``simultaneous'' Nash condition of the estimator is symmetric in $\widehat{\theta}_n$ and $\widehat{\lambda}_n$, unlike the `sequential' mini-max population objective, which nests a family of inner maximizations:
\begin{equation}\label{def:Mestimand}
\lambda_*^\theta = \arg\max_{\lambda\in\Lambda} \mathbb{E} l(\theta, \lambda, Y)
\end{equation}
where the loss $l$ and as a result the solutions $\lambda_*^\theta$ are indexed by the parameter $\theta\in\Theta$. The reader may therefore wonder if we could define an A-estimator for $\theta_*$ in a similar `sequential' mini-max fashion. That is, we could consider a family of M-estimators $\widehat{\lambda}_{n}^\theta$ approximately maximizing the empirical loss at any value of $\theta\in\Theta$:
\begin{equation}
\label{def:Mestimator}
    \widehat{\lambda}_{n}^\theta\in\Lambda_n:\ \ \mathbb{E}_{n}l \left(\theta, \widehat{\lambda}_{n}^\theta,Y \right) \geq \sup_{\lambda \in \Lambda_{n}} \mathbb{E}_{n}l(\theta, \lambda, Y)-\eta_{n}\ \ \forall \theta\in\Theta_n
\end{equation}
And then look for $\widehat{\theta}_n\in\Theta_n$ satisfying: 
\begin{equation}
    \label{def:Aestimator}
    \mathbb{E}_n l(\widehat{\theta}_n, \widehat{\lambda}_n^{\widehat{\theta}_n}, Y) \leq \inf_{\theta\in\Theta_n} \mathbb{E}_n l(\theta, \widehat{\lambda}_n^{\theta}, Y) + \Bar \eta_{n}
\end{equation}
where $\Bar\eta_n=o_\mathbb{P}(1)$ again accommodates approximate minimization.
Fortunately, it turns out that any $\widehat\theta_n$ satisfying the more compact Nash condition from the introduction always satisfies the mini-max condition presented above, as summarized by the following Lemma:
\begin{lemma}
Any $\widehat\theta_n$, satisfying \ref{assn:outernash} and \ref{assn:innernash} for some $\widehat\lambda_n$, also satisfies \ref{def:Aestimator} with some $\widehat\lambda_n^\theta$ for which \ref{def:Mestimator}, $\widehat\lambda_n^{\widehat\theta_n}=\widehat\lambda_n$ and $\Bar\eta_n=\widetilde\eta_n + \eta_n$ holds.
\end{lemma}
\begin{proof}
Pick any $\widehat\theta_n$ satisfying \ref{assn:outernash} and \ref{assn:innernash} for some $\widehat\lambda_n$. Now pick some arbitrary family $\widehat\lambda_n^\theta$ satisfying \ref{def:Mestimator} for all $\theta\neq\widehat\theta_n$, and define $\widehat \lambda_n^{\widehat\theta_n}:=\widehat\lambda_n$. Note that \ref{assn:innernash} directly implies that this $\widehat\lambda_n^\theta$ also satisfies \ref{def:Mestimator} at $\theta=\widehat\theta_n$. It remains to show that the resulting $\widehat\theta_n$ and $\widehat\lambda_n^\theta$ satisfy \ref{def:Aestimator}:
\begin{align*}
    \mathbb{E}_n l(\widehat\theta_n, \widehat\lambda_n^{\widehat\theta_n}, Y) \leq \inf_{\theta\in\Theta_n} \mathbb{E}_n l(\theta, \widehat\lambda_n, Y) + \widetilde\eta_n \leq \inf_{\theta\in\Theta_n} \mathbb{E}_n l(\theta, \widehat\lambda_n^{\theta}, Y) + \widetilde\eta_n + \eta_n
\end{align*}
where the first inequality used $\widehat \lambda_n^{\widehat\theta_n}$ and the Nash condition \ref{assn:outernash}, and the second used the fact that $\widehat\lambda_n^\theta$ was constructed to satisfy \ref{def:Mestimator}.
\end{proof}
This reassures us that it suffices to find one set of values $\widehat\theta_n,\widehat\lambda_n$ which satisfy the Nash condition from the introduction, rather than a continuum of solutions $\widehat\lambda_n^\theta$ indexed by $\theta$. The final $\widehat\theta_n$ will satisfy the mini-max condition regardless, for some (unknown) $\widehat\lambda_n^\theta$. For our theory, it was crucial to derive the uniform convergence of $\widehat\lambda_n^\theta$, hence we will state the rate results for the more general mini-max definition. For the normality result, it was more convenient to work with the stronger Nash definition.

\subsection{Adversaries are Neyman-Orthogonal}
\label{subsec:neyman}
For many A-estimators, one could construct non-adversarial estimators which capture the same population objective. Whenever the adversarial nuisance parameter $\lambda$ is a function, this usually requires a non-parametric first-step estimation of an alternative nuisance parameter. However, such an alternative estimator may not have a desirable property that is guaranteed for A-estimators: {\it Neyman-orthogonality} of $\theta_*$ with respect to the nuisance parameter.\\ 

This property has a long history in statistics, dating back at least to \cite{neyman1959}. It was popularized in econometrics by \cite{Chernozhukov2017DoubleDebiasedNeymanML} as a key setting in which standard machine learning methods can be applied without invalidating causal inference, which sparked follow-up work such as \cite{Chernozhukov2021AutomaticDM} seeking to reformulate non-orthogonal problems as orthogonal ones. The notion applies to parameters which are identified by a moment restriction of the form:
\begin{equation}\label{eq:momentrestriction}
    \mathbb{E}[\varphi(\theta, \nu_*, Y)]=0 \iff \theta=\theta_*
\end{equation}
where $\varphi$ is known and $\nu_*$ is an unknown nuisance parameter which has to be estimated in a first step. A popular estimator $\widehat{\theta}_n$ in this setting would be \cite{hansen1982gmm}'s GMM, for example. The moment condition above is called (Neyman-)orthogonal whenever:
\begin{equation}\label{eq:neyman}
    \nabla_{\nu_*\to\nu}\mathbb{E}[\varphi(\theta_*, \nu_*, Y)]=0 \quad \forall \nu
\end{equation}
Intuitively, this states that the condition identifying $\theta_*$ is ``locally robust'' against perturbations in $\nu_*$. This guarantees that the uncertainty introduced by an appropriate first-step estimation of $\nu_*$ has no first-order effect on the GMM estimator $\widehat{\theta}_n$. Specifically, the asymptotic distribution of $\widehat{\theta}_n$ is the same as in the case in which $\nu_*$ is known. In contrast, when moment restrictions do not satisfy this orthogonality condition, uncertainty about $\nu_*$ generally amplifies the asymptotic variance of $\widehat{\theta}_n$, see e.g. \cite{chenliao2015}, and normality may break down altogether.\\

Notably, if (and only if) $\theta_*$ is parametric, we can examine the first order condition for $\theta_*$ that is implied by the A-estimation objective \ref{def:Aestimand} in this moment restriction framework\footnote{Note however that even when $\theta_*$ is parametric, we usually cannot estimate it via GMM as $\nabla_\theta l(\theta, \widehat\lambda_n^\theta, Y)$ will not exist if $\widehat\lambda_n^\theta$ is a typical sieve, such as a neural network. For the same reason, the theory developed in this paper must not rely on any finite-sample first order conditions. Instead, it will use only the approximate Nash condition \ref{assn:innernash}, \ref{assn:outernash}.} : let $\varphi(\theta, \nu_*, Y)=\nabla_\theta l(\theta, \nu_*(\theta), Y)$, where $\nu_* : \Theta \mapsto \Lambda$ denotes the functional evaluating to $\nu_*(\theta)=\lambda_*^\theta$. Orthogonality then follows from the continuum of first-order conditions identifying $\nu_*$:
\begin{equation*}
\nabla_{ \nu_*\to \nu} \mathbb{E}[l(\cdot, \nu_*(\cdot), Y)]\equiv\mathbf{0} \implies \nabla_{\nu_*\to\nu}\mathbb{E}[\varphi(\theta_*, \nu_*, Y)]=\nabla_{\nu_*\to\nu}\nabla_{\theta_*}\mathbb{E}[l(\theta_*, \nu_*(\theta_*), Y)]=\nabla_{\theta_*} \mathbf{0}
\end{equation*}
since the derivative operators are exchangeable. This implies that as $\widehat\theta_n$ approaches $\theta_*$, an A-estimator $\widehat\theta_n$ is robust to estimation errors in the adversary $\widehat\lambda_n^\theta$ relative to $\lambda_*^\theta$, meaning they do not reduce the accuracy of $\widehat\theta_n$, to a first-order.\\

Consider the example of Section \ref{ex:fDiv}, which estimates $\theta_*$ minimizing the f-Divergence between the model $\mathbb{P}_\theta$ and the data $\mathbb{P}$. As a non-adversarial alternative, we could re-parametrize the problem and estimate $\nu_*:=d\mathbb{P}$ via a first-step Kernel density estimator $\widehat{\nu}_n(Y)=\widehat{d\mathbb{P}}_n(Y)$, and subsequently approximate the f-Divergence as the average over $f\left(\frac{d\mathbb{P}_\theta(Y)}{\widehat{\nu}_n(Y)}\right)$. However, the first-order condition for $\theta_*$ would not satisfy orthogonality, hence a GMM estimator based on this condition may not attain the variance of the analogous GMM estimator using $\nu_*$ instead. In contrast, the A-estimator of Section \ref{ex:fDiv} {\it does} attain this variance - due to its orthogonal adversary - which we formally establish in Section \ref{app:fDiv}. Moreover, this remains true when generalizing to a setting in which $\theta_*$ contains unknown functions, where no analogous GMM estimator exists that could capture the continuum of first-order conditions in $\theta_*$.\looseness=-1

\subsection{Convergence Rate of A-Estimators}
\label{subsec:generalrate}
We begin with a general theorem characterizing the convergence rates of sieve A-estimators, for arbitrary loss functions and parameter spaces. It can be viewed as a generalization of \cite{shen1994}'s M-estimator result. Its proof is provided in Appendix \ref{proof:convergence}, with the main challenge being that \cite{shen1994}'s chaining arguments need to be carefully modified to hold uniformly over $\Theta$. Our theorem adopts a more compact formulation than \cite{shen1994} which does not require any norm over $\Theta,\Lambda$ to state our assumptions, although convergence rates are obtained for any (pseudo-)norm $d(\theta,\theta_*)$ which is dominated by the objective.
\begin{theorem}[Convergence Rate of A-Estimators]\label{thm:generalrate} Assume that:
\begin{itemize}[leftmargin=*]
    \item C1: The criterion variance is bounded by a power $\gamma > 0$ of its expectation:
    \begin{align}
        \mathbb{V}[l(\theta, Y)-l(\theta_*, Y)] & \prec \mathbb{E}[l(\theta, Y)-l(\theta_*, Y)]^{\gamma} \label{assn:aloss}\\
        \mathbb{V}[l^\theta(\lambda_*^\theta, Y)-l^\theta(\lambda, Y)] & \prec \mathbb{E}[l^\theta(\lambda_*^\theta, Y)-l^\theta(\lambda, Y)]^\gamma \label{assn:mloss}
    \end{align}
    for all $\theta\in\Theta,\lambda\in\Lambda$ for which the right hand sides are less than some constant.
    \item C2: For all small $\varepsilon>0$, the covering number (Def. \ref{def:coveringnumber}) is bounded via \begin{align}
        \log \mathcal{N}(\varepsilon, \{l(\theta, \lambda, \cdot): \theta\in\Theta_n, \lambda\in\Lambda_n\}, \|\cdot\|_\infty) & \prec n^{s}(\varepsilon^{-r}-1)/r \label{eq:covassn}
    \end{align}
    for $0\leq s<1$ and $r\geq 0$, where $r=0$ represents $\lim_{r\to0}n^{s}(\varepsilon^{-r}-1)/r=n^{s}\log(1/\varepsilon)$.
\end{itemize}
Then the following conclusions hold.
\begin{itemize}[leftmargin=*]
    \item[i)] The criterion converges at rate: \begin{align}
    \mathbb{E}[l(\theta_*, Y)-l(\widehat{\theta}_n, Y)] & =O_\mathbb{P}(n^{-\tau(\gamma, s, r, n)}  + \epsilon_n+\eta_n+ \Bar \epsilon_n+\Bar \eta_n) \label{eq:aresult} \\
    \sup_{\theta\in\Theta_n} \mathbb{E}[l^\theta(\lambda_*^\theta, Y)-l^\theta(\widehat{\lambda}_n^\theta, Y)] & =O_\mathbb{P}(n^{-\tau(\gamma, s, r, n)} + \epsilon_n+\eta_n) \label{eq:mresult}
\end{align}
where $\Bar\epsilon_n = \mathbb{E}[l(\pi_n\theta_*, Y)-l(\theta_*, Y)]$ and $\epsilon_n = \sup_{\theta\in\Theta_n} \mathbb{E}[l^\theta(\lambda_*^\theta, Y)-l^\theta(\pi_n \lambda_*^\theta, Y)]$ are the sieve approximation errors. \ref{eq:mresult} also holds without \ref{assn:aloss}. $\tau(\gamma, s, r, n)$ represents:
$$
\tau(\gamma, s, r, n)=\left\{\begin{array}{ll}
{1-s}-\frac{\log \log n}{\log n}, & \text { if } r=0, \gamma \geq 1 \\
\frac{1-s}{2-\gamma}, & \text { if } r=0, \gamma<1 \\
\frac{1-s}{2-\min (1, \gamma)(2-r)/2}, & \text { if } 0<r<2 \\
\frac{1-s}{2}-\frac{\log \log n}{\log n}, & \text { if } r=2 \\
\frac{1-s}{r}, & \text { if } r>2
\end{array}\right.
$$
\item[ii)] Hence, $d(\widehat\theta_n, \theta_*)=o_\mathbb{P}(1)$ for any (pseudo-)norm $d(\cdot,\cdot)$ under which $\mathbb{E}[l(\theta, Y)]$ compact and continuous. If also $d(\theta, \theta_*)^{1/q}\prec \mathbb{E}[l(\theta, Y)-l(\theta_*,Y)]$ for $q>0$, we get:
$$d(\widehat\theta_n, \theta_*)=O_\mathbb{P}(n^{-\tau(\gamma, s, r, n)q}  + \epsilon_n^{q}+\eta_n^{q}+ \Bar \epsilon_n^{q}+\Bar \eta_n^{q})$$
\end{itemize}
\end{theorem}
\begin{remark}[Discussion of Assumptions] 
The theorem extends \cite{shen1994}'s convergence rate result for sieve M-estimators to A-estimators. There is a direct mapping between our assumptions and theirs: our C1 combines their assumptions C1 and C2, and our C2 corresponds to their C3. Our proof in Appendix \ref{proof:convergence} is structured in the same way as that of \cite{shen1994}, although we need to modify their Lemmas to obtain the uniform convergence of the adversary in \ref{eq:mresult}, which is crucial to the main result \ref{eq:aresult}. The key modifications to our assumptions, which allow us to do so are: C1) that the constant factor implicit in the ``$\prec$'' relation of \ref{assn:mloss} must not depend on $\theta$, as implied by the definition of ``$\prec$'' at the beginning of this section and C2) that the complexity of the {\it joint} sieve space $\Theta_n\times\Lambda_n$ satisfies the entropy bound. Otherwise, the assumptions are conceptually the same and we refer the reader to \cite{shen1994} for a more detailed discussion.
\end{remark}

\begin{remark}
Using similar arguments as ours, one may establish the uniform convergence of A-estimators over a third parameter space, generalizing the setting to arbitrary finite sequences of $\min$'s and $\max$'s over different parameter spaces: e.g. $\min_\theta\max_\lambda\min_\gamma\mathbb{E}[l(\theta,\lambda,\gamma,Y)]$. This would yield convergence rates towards more general {\it Stackelberg equilibria} in so-called {\it empirical games}, for which we are currently only aware of a consistency result by \cite{tuyls2018generalised}.
\end{remark}
\begin{remark}
Beyond convergence rates for $\widehat{\theta}_n$ and $\widehat{\lambda}_n^\theta$, it is often useful to control the empirical process of arbitrary functions $f(\theta, \lambda, Y)$ of the parameters, e.g. to establish conditions for asymptotic normality required by Theorem \ref{thm:normality}. For this purpose, we provide Lemma \ref{lem:emproc} in Appendix \ref{sec:lemmas}.
\end{remark}

\subsection{Semiparametric Rates with Neural Networks}
\label{subsec:neuralArate}
Next, we will apply the general result of the previous section to derive the convergence rates for neural network A-estimators. For generality, we will consider the semiparametric setting in which $\theta,\lambda$ may contain both Euclidean vectors and functions. These lower-level conditions are easy to verify in practice, but are general enough to apply to all estimators considered in Section \ref{sec:examples}. We will include the proof as it is short and an instructive application of Theorem \ref{thm:generalrate}. The theorem allows for two types of function classes, both of which can be viewed as generalizations of traditional Hölder functions with $D$-dimensional domain, with their own notion of an {\it intrinsic dimension} $d^*\leq D$, which may be smaller than that of the data $D$. As we will review in Remark \ref{rem:intrinsic}, we observe that neural networks achieve a reduced curse of dimensionality in these settings.\looseness=-1
\begin{theorem}[Semiparametric Rates with Neural Networks]\label{thm:neuralArate} \ \\
Consider the semiparametric setting in which $\Theta=\Bar{\mathcal{B}}\times\Bar{\mathcal{A}}$ and $\Lambda=\mathcal{B}\times\mathcal{A}$, where $\Bar{\mathcal{B}},\mathcal{B}$ are subsets of some Euclidean spaces and $\Bar{\mathcal{A}}, \mathcal{A}$ are some function spaces. Let $\Lambda, \Theta$ be compact under $\|\cdot\|_\infty$. For all $\lambda,\lambda'\in\Lambda,\ \theta,\theta'\in\Theta$, assume the following conditions hold:
\begin{itemize}[leftmargin=*]
    \item A0: Assume that $\theta_*\in \Theta_*$ satisfies either\begin{itemize}
        \item[a)] $\Theta_*\subset\Bar{\mathcal{B}}\times\mathcal{H}(\Bar p, \Bar {\mathcal{X}})$ on some $ \Bar {\mathcal{X}}\subset[0,1]^{\Bar D}$ with $\operatorname{dim}_M \Bar {\mathcal{X}} = \Bar d^* \leq \Bar D $ (see Def. \ref{def:minkowski} and \ref{def:holder})
        \item[b)] $\Theta_*\subset\Bar{\mathcal{B}}\times\mathcal{G}( \Bar p ,\Bar d^* , [0, 1]^{\Bar D })$ (see Def. \ref{def:GHI}) 
    \end{itemize}
    and that $\{\lambda_*^\theta : \theta\in\Theta\}\subset\Lambda_*$ satisfies either \begin{itemize}
        \item[a)] $\Lambda_*\subset\mathcal{B}\times\mathcal{H}(p , {\mathcal{X}} )$ on some ${\mathcal{X}} \subset[0,1]^{D }$ with $\operatorname{dim}_M {\mathcal{X}}  = d^* \leq D $ 
        \item[b)] $\Lambda_*\subset\mathcal{B}\times\mathcal{G}(p , d^* , [0, 1]^{D })$ 
    \end{itemize}
    \item A1: $l(\theta, \lambda, Y)-l(\theta', \lambda', Y) \prec \|\theta-\theta' \|_{\Bar {\mathcal{X}} } + \|\lambda-\lambda'\|_{{\mathcal{X}} }$
    \item A2:  $\mathbb{V}[l(\theta, Y)-l(\theta_*, Y)] \prec \mathbb{E}[l(\theta, Y)-l(\theta_*, Y)] \prec \|\theta-\theta_*\|^{2}_{\widetilde{{\mathcal{X}}}} + \mathbb{P}(\Bar x\not\in\widetilde{{\mathcal{X}}}) \ \forall \widetilde{{\mathcal{X}}}\subset\Bar {\mathcal{X}}$
    \item A3: $\mathbb{V}[l^\theta(\lambda_*^\theta, Y)-l^\theta(\lambda, Y)] \prec \mathbb{E}[l^\theta(\lambda_*^\theta, Y)-l^\theta(\lambda, Y)]\prec \|\lambda-\lambda_*^\theta\|^{2}_{\widetilde{{\mathcal{X}}}} + \mathbb{P}(x\not\in \widetilde{\mathcal{X}}) \ \forall \widetilde{{\mathcal{X}}}\subset {\mathcal{X}}$
\end{itemize}
Pick any two values $\Bar r>\underline r\geq \left(\frac{d^*}{p}\vee \frac{\Bar d^*}{\Bar p}\right)$. Consider the A-estimator \ref{def:Mestimator} with $\eta_n, \Bar\eta_n = o_\mathbb{P}(n^{-2/(2+ \Bar r)})$ where $\Lambda_n=\mathcal{B}\times\mathcal{F}_{\sigma}(L, W_n, w_n, \kappa_n)$ and $\Theta_n=\Bar{\mathcal{B}}\times\mathcal{F}_{\sigma}(\Bar L, \Bar W_n, \Bar w_n, \Bar \kappa_n)$ implement neural networks (cf. Definition \ref{def:neuralnets}) satisfying $W_n, \Bar W_n, w_n, \Bar w_n \asymp n^{\underline r/(\underline r+2)}$ and $\kappa_n, \Bar \kappa_n \asymp n^{c}$ for any large enough choice of $L,\Bar L, c>0$. For A0a) choose $\sigma(x)=\mathrm{ReLU}(x)$ and for A0b) choose $\sigma(x)=\tanh(x)$. Then:
\[\mathbb{E}[l(\widehat{\theta}_n, Y)-l(\theta_*, Y)]=o_{\mathbb{P}}(n^{-2/(2+\Bar r)})\]
\[\sup_{\theta\in\Theta_n} \mathbb{E}[l^\theta(\lambda_*^\theta, Y)-l^\theta(\widehat\lambda_n^\theta, Y)]=o_{\mathbb{P}}(n^{-2/(2+\Bar r)})\]
Hence, $d(\widehat\theta_n, \theta_*)=o_\mathbb{P}(1)$ for any (pseudo-)norm $d(\cdot,\cdot)$ under which $\mathbb{E}[l(\theta, Y)]$ is compact and continuous. Further, if $d(\theta, \theta_*)^{1/q}\prec \mathbb{E}[l(\theta, Y)-l(\theta_*,Y)]$ for $q>0$, we get:\looseness=-1
$$d(\widehat\theta_n, \theta_*)=o_\mathbb{P}(n^{-2q/(2+\Bar r)})$$
\end{theorem}
\begin{proof}\singlespacing
We will verify the conditions of Theorem \ref{thm:generalrate}. A2 and A3 imply C1 (\ref{assn:aloss} and \ref{assn:mloss}) with $\gamma=1$. Lipschitzness A1 together with Lemma \ref{lem:neuralentropy} imply C2 (\ref{assn:entropy}) with $s=t/(t+2)$ for any $t: \Bar r>t> \underline{r}$ and $r=0$. Therefore Theorem \ref{thm:generalrate} applies with $n^{-\tau(\gamma, s, r, n)}=n^{2/(2+t)}\log n\prec n^{2/(2+\Bar r)}$, which dominates $\eta_n$ and $\Bar{\eta}_n$ by assumption. We are therefore left with bounding $\epsilon_n$ and $\Bar \epsilon_n$. By A3, we can bound $\epsilon_n \prec \sup_{\theta\in\Theta_n} \|\pi_n \lambda_*^\theta-\lambda_*^\theta\|_{\widetilde{{\mathcal{X}}}}^2 + \mathbb{P}(x\not\in\widetilde{{\mathcal{X}}})$ for any $\widetilde{{\mathcal{X}}}\subset {\mathcal{X}}$. In the case of A0a), we set $\widetilde{{\mathcal{X}}}={\mathcal{X}}$ and use Lemma \ref{lem:approxminkowski} to obtain $\sup_{\theta\in\Theta_n} \|\pi_n \lambda_*^\theta-\lambda_*^\theta\|_{\mathcal{X}}^2 \prec (W_n \wedge w_n)^{-2 p/d^*}\prec n^{-2p\underline r/d^*/(2+\underline r)}\prec n^{2/(2+\underline r)}$ which yields $\epsilon_n = o(n^{2/(2+ \Bar r)})$. For A0b), Lemma \ref{lem:approxGHI} yields the same bound as Lemma \ref{lem:approxminkowski}, but only over a subset $\widetilde{{\mathcal{X}}}\subset {\mathcal{X}}$ with $P(x\not\in \widetilde{{\mathcal{X}}})\prec n^{-k}$ for some arbitrarily large constant $k>0$, which only affects the constant $c$ in the bound on $\kappa_n$. Hence we conclude that $\epsilon_n \prec n^{2/(2+ \underline r)} + n^{-k} \prec n^{2/(2+ \underline r)}$. Analogous arguments yield the same bound for $\Bar \epsilon_n$.
\end{proof}\onehalfspacing
\begin{remark}[Discussion of Assumptions] A0 defines the function classes addressed by the Theorem. Both are generalizations of traditional Hölder classes which arise for $d^*=D$, see Remark \ref{rem:intrinsic}. Condition A1 requires the loss to be Lipschitz in both parameters, which simplifies (but is not necessary for) the verification of C2. Condition A2 (and analogously A3) consists of two parts. First, it states that for a given parameter, the variance of the criterion difference must be bounded by its expectation, a simplified version of Assumption C1 of Theorem \ref{thm:generalrate} which happens to be satisfied in all of our examples, but versions of this Theorem with $\gamma\neq 1$ can be derived via the same steps as the proof above. The second part of the condition bounds the expected loss by a squared $\sup$-norm over any subset $\widetilde{\mathcal{X}}$ of the function domain $\mathcal{X}$. For the case of A0a), it would have sufficed to state the condition with $\widetilde{\mathcal{X}}=\mathcal{X}$ only, but for A0b) we require arbitrary subsets $\widetilde{\mathcal{X}}$ to apply the approximation result of Lemma \ref{lem:approxGHI}. A2 is implied, for example, by $\mathbb{E}[l(\theta, Y)-l(\theta_*)]\prec \|h(\theta)-h(\theta*)\|^2_{\mathcal{L}^q(\mathcal{X})}$ for some $q$ and Lipschitz map $h: \Theta\mapsto\Theta$. The assumption is significantly weaker than \cite{shen2019asymptotic} or \cite{farrell2018deep} who impose $\mathbb{E}[l(\theta, Y)-l(\theta_*)]\asymp \|\theta-\theta*\|^2_{\mathcal{L}^2(\mathcal{X})}$, which would not hold for Examples \ref{ex:GEL} or \ref{ex:CMR}. It could be generalized further to allow for arbitrary powers of the $\sup$-norms (and proved in the same way via Theorem \ref{thm:generalrate}), but the squares arise rather universally via Taylor expansions.
\end{remark}

\begin{remark}\label{rem:intrinsic}
Theorem \ref{thm:neuralArate} clarifies that neural networks do not necessarily exhibit the curse of dimensionality, as the lower bound on $\Bar r$ does not depend on the dimension $D$ of the data. Instead, what matters is the {\it intrinsic dimension} $d^*$ of the target function. In the setting A0a), introduced by \cite{minkowski}, $d^*$ refers to the Minkowski dimension of the manifold ${\mathcal{X}}$ which supports the data. It has been observed that $d^*\ll D$ for many high-dimensional types of data: intuitively, $d^*$ is low whenever there is strong statistical dependency between the individual dimensions of the data. Examples include the characteristics of physical products, images and natural language. In the setting A0b), introduced by \cite{bauer2019deep}, $d^*$ refers to the order of a generalized hierarchical interaction model. It is common for structural models in e.g. economics or optimal control to suggest that an unknown function is hierarchically composed of some finite number of individual functions which only depend on $d^*\ll D$ inputs at a time. The result underscores that neural networks can {\it adaptively} - that is, without the researcher modifying the estimation procedure - exploit {\it structures} in the target function which allow them to model the relationships more efficiently than what standard convergence results suggest.\end{remark}

\subsection{Asymptotic Normality of A-Estimators}
\label{subsec:normality}
In applications, it we a often interested in estimating a quantity of the form $F(\theta_*)$, where $F:\Theta\mapsto\mathbb{R}$ is some known functional. To derive confidence intervals around the plug-in estimate $F(\widehat\theta_n)$, we need its asymptotic distribution. To this end, we present Theorem \ref{thm:normality}, which can roughly be viewed as a generalization of \cite{shen1997onmethods} to A-Estimators. For this section, we make use of the pathwise derivative presented in Definition \ref{def:derivatives}. We require a particular inner product over the space $\Theta$:
\[\langle \theta, \theta' \rangle := \nabla_{\theta_*\to\theta} \nabla_{\theta_*\to\theta'} \mathbb{E}[l(\theta_*, Y)]\]
As discussed in Definition \ref{def:derivatives}, the notation $\nabla_{\theta_*\to\theta}$ implicitly assumes that the corresponding limit exists and is linear in $\theta$. For short, we write $\lambda_*^{\prime\theta}[v]:=\nabla_{\theta\to v}\lambda_*^\theta$, $l'(\theta, Y)[v]:=\nabla_{\theta\to v}l(\theta, Y)$ and $l'(\theta,\lambda, Y)[v, w]:=\nabla_{\theta\to v} l(\theta,\lambda, Y) + \nabla_{\lambda\to w} l(\theta, \lambda, Y)$.
\begin{theorem}[General Normality of A-Estimators]\label{thm:normality}\ \\
Consider the estimators $\widehat\theta_n, \widehat\lambda_n$ satisfying the Nash conditions \ref{assn:outernash} and \ref{assn:innernash}. Fix a sequence $e_n=o(n^{-1/2})$. Assume $F$ is smooth enough and $\widehat\theta_n, \widehat\lambda_n$ converge fast enough such that a Riesz representer $v_*\in\Theta_*$ exists, satisfying:
\begin{equation}
\label{assn:smoothf}
    \sup_{\theta\in\widehat{\Theta}_n} |F(\theta)-F(\theta_*)-\langle \theta -\theta_*, v_*\rangle| = O_\mathbb{P}(e_n)
\end{equation}
Where $\widehat{\Theta}_n$ and $\widehat{\Lambda}_n(\theta)$ are the shrinking neighborhoods defined in Lemma \ref{lem:emproc}. For $v\in\{v_*, -v_*\}$, define the local perturbations $\Bar{\theta}_n(\theta)=\theta-e_n v$ and $\Bar{\lambda}_n^\theta(\lambda)=\lambda + e_n \lambda^{\prime\theta}_*[v]$ and assume:

\begin{itemize}
\item[] CONDITION N1: Stochastic Equicontinuity
$$
\sup_{\theta\in\widehat{\Theta}_n,\lambda\in\widehat{\Lambda}_n(\theta)} (\mathbb{E}_n-\mathbb{E}) l'(\theta,\lambda, Y)[v,\lambda_*^{\prime\theta}[v]]-l'(\theta_*, Y)[v] = O_\mathbb{P}(e_n)
$$
\item[] CONDITION N2: Population Criterion Difference
$$
\sup_{\theta\in\widehat{\Theta}_n,\lambda\in\widehat{\Lambda}_n(\theta)} \mathbb{E} l'(\theta,\lambda, Y)[v,\lambda_*^{\prime\theta}[v]]-l'(\theta_*, Y)[v] - \langle \theta-\theta_*, v \rangle =  O_\mathbb{P}(e_n) 
$$
\item[] CONDITION N3: Approximation Error
$$
\sup_{\theta\in\widehat{\Theta}_n,\lambda\in\widehat{\Lambda}_n(\theta)} \mathbb{E}_n l'(\theta,\lambda,Y)[\Bar{\theta}_n(\theta)-\pi_n\Bar{\theta}_n(\theta), \Bar{\lambda}_n^\theta(\lambda)-\pi_n\Bar{\lambda}_n^\theta(\lambda)]  = O_\mathbb{P}(e_n^2)
$$

\end{itemize}

If \ref{assn:outernash} and \ref{assn:innernash} are satisfied with $\widetilde\eta_n, \eta_n = O_\mathbb{P}(e_n^2)$, then:
\[\sqrt{n} \left( F(\widehat{\theta}_n)-F(\theta_*)\right) \overset{d}{\longrightarrow} \mathcal{N}(0, V),\ \text{ where }V=\mathbb{V}\left(l'(\theta_*, Y)[v_*]\right)\]
\end{theorem}
\begin{remark}[Discussion of Assumptions]
In contrast to our convergence rate result, our proof requires the A-estimator to satisfy the (stronger) Nash condition from the introduction. Our conditions N1-3 are analogues of \cite{shen1997onmethods}'s and play the same roles in our proof. N1 combines their assumptions A and D, N2 corresponds to their B, and N3 to their C. \cite{shen1997onmethods}'s high-level discussion of their assumptions therefore applies to ours as well, and we again refer the reader there for additional context. The main difference is that their conditions are formulated to control the remainder of a second order Taylor expansion, whereas we look at the convergence of the first derivative, which results in $O_\mathbb{P}(e_n)=o_\mathbb{P}(n^{-1/2})$ requirements for N1 and N2, rather than the $O_\mathbb{P}(e_n^2)=o_\mathbb{P}(n^{-1})$ found in \cite{shen1997onmethods}'s conditions A and B. 
\end{remark}
\begin{remark}\label{rem:nonconvex}
Condition N3 is a version of a known condition on approximation error in M-estimation settings (see Condition C4 in \cite{shen2019asymptotic} and Condition C in \cite{shen1994}). Its verification usually exploits convexity of $\Theta_n$, such that $\pi_n\Bar{\theta}_n(\theta)=\theta+e_n\pi_n v_*$. This holds for series or kernel based estimators, but not neural networks. \cite{shen2019asymptotic} therefore leave it as an explicit assumption, concluding that it is unclear how to verify it for neural networks. In Theorem \ref{thm:neuralAnormality}, we resolve this issue, showing that N3 can be verified for non-convex sieves such as neural networks by adhering to two simple implementation choices: 1) \textit{undersmoothing}, i.e. choosing a sieve which grows faster than rate-optimal, achieving an approximation error of $o(n^{-1})$ and 2) regularizing the sieves towards the convex target classes containing $\theta_*,\lambda_*$.\looseness=-1
\end{remark}

\subsection{Semiparametric Normality with Neural Networks}
\label{subsec:neuralAnormality}
Next, we present Theorem \ref{thm:neuralAnormality}, which strengthens the assumptions of our previous neural network convergence rate result (Theorem \ref{thm:neuralArate}) in a way that allows us to derive the asymptotic normality of functionals $F(\widehat\theta_n)$ via Theorem \ref{thm:normality}. A crucial innovation is that we are able to work around the non-convexity issues of deep neural networks discussed in Remark \ref{rem:nonconvex}, to obtain a normality result from {\it low-level} conditions, which only consist of general properties that the loss function must satisfy (A4-A7), and certain {\it implementation choices} for the neural networks that must be followed. To the best of our knowledge, the theorem therefore also provides the first low-level conditions for the normality of smooth functionals of deep neural network {\it M-estimators} (as the special case where $\Lambda$ is singleton). 

\begin{theorem}[Semiparametric Normality with Neural Networks]\label{thm:neuralAnormality}\ \\
Let all assumptions of Theorem \ref{thm:neuralArate} be satisfied with $\frac{d^*}{p}\vee\frac{\Bar{d}^*}{\Bar{p}}<1/4$, and choose $2\geq\Bar r>\underline r> 2/3$. Let $\Theta_*, \Lambda_*$ be convex and $\theta_*, \lambda_*^\theta$ lie in their interior. Replace the neural network sieves $\Theta_n,\Lambda_n$ with the following regularized versions:
\[\Theta_n \gets \{\theta \in\Theta_n: \inf_{\theta'\in\Theta_*} \|\theta-\theta'\|_{\Bar {\mathcal{X}}} \prec n^{-1-\epsilon}\},\quad \Lambda_n \gets \{\lambda \in\Lambda_n: \inf_{\lambda'\in\Lambda_*} \|\lambda-\lambda'\|_{{\mathcal{X}}} \prec n^{-1-\epsilon}\}\]
for any $\epsilon>0$ which is small enough to guarantee that $\Theta_n, ,\Lambda_n$ are nonempty. Further, for all $\theta, v\in\Theta,\ \lambda, w\in\Lambda$, assume:
\begin{itemize}[leftmargin=*]
    \item A4: Lipschitz Derivative: $l'(\theta, \lambda, Y)[v,w]-l'(\theta', \lambda', Y)[v,w] \prec \|\theta-\theta'\|_{\Bar {\mathcal{X}}}+\|\lambda-\lambda'\|_{\mathcal{X}}$
    \item A5: The perturbations are smooth: $v_*\in\Theta_*$, $\lambda_*^{\prime\theta}[v_*]\in\Lambda_*$
    \item A6: The Taylor remainders vanish with the loss:
    \begin{itemize}
    \item[i)] $|\mathbb{E}l'(\theta, \lambda, Y)[v_*,\lambda_*^{\prime\theta}[v_*]]-l'(\theta, Y)[v_*]| \prec \mathbb{E}[l(\theta, \lambda_*^\theta, Y)-l(\theta, \lambda, Y)]$
    \item[ii)] $|\mathbb{E}l'(\theta, Y)[v_*]-l'(\theta_*, Y)[v_*]-\langle \theta-\theta_*, v_*\rangle|\prec \mathbb{E}[l(\theta, Y)-l(\theta_*, Y)]$
    \end{itemize}
    \item A7: For non-Donsker classes, the variance of the derivatives is bounded by the loss:
    \begin{itemize}
    \item[i)] $\mathbb{V}[l'(\theta, \lambda, Y)[v, \lambda_*^{\prime\theta}[v]] - l'(\theta, \lambda_*^{\theta}, Y)[v, \lambda_*^{\prime\theta}[v]]] \prec \mathbb{E}[l(\theta, \lambda_*^\theta, Y)-l(\theta, \lambda, Y)]$ or $\Lambda_*$ is $\mathbb{P}$-Donsker
    \item[ii)] $\mathbb{V}[l'(\theta, Y)[v] - l'(\theta_*, Y)[v]] \prec \mathbb{E}[l(\theta, Y)-l(\theta_*, Y)]$ or $\Theta_*$ is $\mathbb{P}$-Donsker
    \end{itemize}
\end{itemize}
If $\widehat{\theta}_n,\widehat{\lambda}_n$ satisfy the Nash condition \ref{assn:outernash},\ref{assn:innernash} with $\eta_n,\Tilde\eta_n=o_\mathbb{P}(n^{-1})$, then:
\[\sqrt{n} \left( F(\widehat{\theta}_n)-F(\theta_*)\right) \overset{d}{\longrightarrow} \mathcal{N}(0, V),\ \text{ where }V=\mathbb{V}\left(l'(\theta_*, Y)[v_*]\right)\]
\end{theorem}
\begin{remark}[Discussion of Assumptions]
The Theorem requires that the neural network sieves $\Theta_n,\Lambda_n$ are implemented to undersmooth (i.e. grow faster than the rate-optimal sieve would) via the condition on $\underline{r}$, while being regularized towards the convex target spaces $\Theta_*, \Lambda_*$. Note that this does not affect the sieve's approximation power towards these spaces, and there always exists an $\epsilon>0$ for which $\Theta_n,\Lambda_n$ are non-empty due to their $o(n^{-1})$ approximation rates. While in principle just an implementation choice, the current $\sup$-norm regularization is arguably not practical and future work may be able to clarify whether e.g. an appropriate L2 penalty on the weights suffices. Conditions A4-A7 are general conditions on the loss function which can be satisfied in all our examples. A4 is a simple Lipschitz condition analogous to A1. The smoothness of the Riesz representer (A5) is most easily verified by computing and examining a given $v_*, \lambda_*^{\prime\theta}[v_*]$ directly, although the Riesz representation theorem can provide general conditions under which $v_*$ lives in the same space as $\theta_*$. A6 is a standard condition controlling the Taylor remainder. For a discussion, see e.g. Assumptions 4.5 in \cite{aichen2003} and \cite{aichen2007}, or Assumption 3.5ii) in \cite{sievewald2015}. Whether it holds depends on how non-linear the objective is: e.g. for the quadratic objective of \cite{dikkala2020minimax}, the left-hand side is zero. A7 serves to control the empirical process (N1). It can be easily satisfied either by bounding the variances of the derivatives, or by relying on the Donsker property of the target space (cf. Remark \ref{rem:donsker}).
\end{remark}
\begin{remark}\label{rem:donsker}
Note that the Donsker property and thus A7 always holds if $p>D/2$, where standard results using bracketing number bounds imply that the Hölder spaces $\Theta_*, \Lambda_*$ satisfy the Donsker property. We conjecture that this analogously holds for our lower-dimensional classes A0a) and A0b) whenever $d^*/p<2$, which would make the verification of A7 unnecessary in general, since we require $\frac{d^*}{p}\vee\frac{\Bar{d}^*}{\Bar{p}}<1/4$. Verifying this conjecture is beyond the scope of this paper however, hence we provide A7 as an explicit assumption for maximum flexibility.
\end{remark}

\section{Application to Examples}
\label{sec:application}

\subsection{Minimum $f$-Divergence} \label{app:fDiv}
Applying our general Theorem \ref{thm:neuralArate} to the estimator of Section \ref{ex:fDiv} yields Proposition \ref{thm:fdivrates}, which provides the convergence rate of semiparametric $\widehat\theta_n$ if $\Lambda$ and all unknown functions in $\Theta$ are approximated by classes of neural networks.
\begin{proposition}\label{thm:fdivrates}
Let $\theta_*\in\Theta_*\subset\Theta,\lambda_*^\theta=f'(\frac{\D\mathbb{P}_\theta}{\D\mathbb{P}})\in\Lambda_*\subset\Lambda$, where $\Theta,\Lambda$ are compact under $\|\cdot\|_\infty$ and path-connected, and the target function classes $\Theta_*,\Lambda_*$ satisfy A0 in Theorem \ref{thm:neuralArate}. Fix some $C<\infty$. For any $\theta\in\Theta$, let $0<f''\left(\frac{\D\mathbb{P}_\theta}{\D\mathbb{P}}(Y)\right)<C$ wp1 and for any $\lambda\in\Lambda$, let $0<f''_*(\lambda(Y))<C$ wp1. Let $\left\|\frac{\D\mathbb{P}_\theta}{\D\mathbb{P}}-\frac{\D\mathbb{P}_{\theta'}}{\D\mathbb{P}}\right\|_\infty\prec \|\theta-\theta'\|_\infty$. Let $\Theta_n, \Lambda_n$ be constructed as in Theorem \ref{thm:neuralArate}, with all neural networks growing in width at some rate $n^{\underline{r}/(\underline{r}+2)}$ satisfying $\underline{r}\geq \frac{d^*}{p} \vee \frac{\Bar d^*}{\Bar p}$. Then for any $\Bar{r}>\underline{r}$:
$$D_f(\mathbb{P}_{\widehat{\theta}_n} \| \mathbb{P}) = o_\mathbb{P}(n^{-2/(2+\Bar{r})})$$
\end{proposition}
\begin{remark}
In general, the convergence rate of $\widehat{\theta}_n$ is faster the slower the growth rate $n^{\underline{r}/(\underline{r}+2)}$ of the neural network. However, the growth must be fast enough to control the approximation error of the sieves $\Theta_n, \Lambda_n$ relative to the target function classes $\Theta_*,\Lambda_*$. This lower bound depends on the ratio of the {\it smoothness} of the target classes $p$ and $\Bar p$ and their {\it intrinsic dimensions} $d^*$ and $\Bar d^*$, which may be smaller than that of the data $Y$, in which case $f$-GANs attain a reduced curse-of-dimensionality relative to traditional nonparametric density estimators.
\end{remark}
\begin{remark}
This convergence rate result stands in contrast to \cite{generalizationGAN}, who argued that Generative Adversarial Networks do not generalize with respect to the metric given by the population objective, only under a weaker ``neural net distance'' which they introduce. The convergence rate result above clarifies that the broad class of $f$-GANs in fact {\it does} converge quickly under population divergence.
\end{remark}

While a fast convergence rate of the model distribution $\mathbb{P}_{\widehat\theta_n}$ is a key goal in semi- and nonparametric estimation, whenever some function $F(\widehat\theta_n)$ of the estimate informs downstream decision-making, we are often interested in obtaining confidence intervals around $F(\widehat\theta_n)$. To this end, we derive the asymptotic normality of the adversarial $f$-Divergence objective - an entirely novel result at this level of generality, to the best of our knowledge. First, we compute the inner product defined in Section \ref{subsec:normality}, which can be expressed concisely:
$$
    \langle \theta, \theta' \rangle = \nabla_{\theta_*\to\theta}\nabla_{\theta_*\to\theta'} \mathbb{E}\left[f\left(\frac{\D\mathbb{P}_{\theta_*}(Y)}{\D\mathbb{P}(Y)}\right)\right] = \mathbb{E}\left[\nabla_{\theta_*\to\theta}\log \D\mathbb{P}_{\theta_*}(Y) \cdot \nabla_{\theta_*\to\theta'}\log \D\mathbb{P}_{\theta_*}(Y)\right]
$$
Where $\nabla_{\theta_*\to\theta}\log \D\mathbb{P}_{\theta_*}(Y)=\nabla_{\theta_*\to\theta}\frac{\D\mathbb{P}_{\theta_*}(Y)}{\D\mathbb{P}(Y)}$ is a pathwise derivative of the Radon-Nikodym derivative. Conditions under which the normality result of Section \ref{subsec:normality} applies are presented in Proposition \ref{thm:fdivnormality}.
\begin{proposition}\label{thm:fdivnormality}
Consider a functional $F(\theta)$ for which a Riesz representer $v_*$ exists satisfying \ref{assn:smoothf} with $\langle\cdot,\cdot\rangle$ defined above. Let all assumptions of Theorem \ref{thm:fdivrates} be satisfied for $d^*/p \vee \Bar d^*/\Bar p < 1/4$ and assume that $\mathcal{\Theta}_*$ is Donsker. Let $\Theta_*, \Lambda_*$ be convex, let $\theta_*, \lambda_*^\theta$ lie in their interior, and let them contain $v_*, \lambda^{\prime,\theta}_*[v_*]$. Assume the Lipschitz condition $\|\nabla_{\theta\to v}\frac{\D\mathbb{P}_\theta}{\D\mathbb{P}}-\nabla_{\theta'\to v}\frac{\D\mathbb{P}_{\theta'}}{\D\mathbb{P}}\|_\infty\prec \|\theta-\theta'\|_\infty$ and let $f''$ be Lipschitz. Pick $2\geq\Bar r >\underline{r}>2/3$ and regularize $\Theta_n, \Lambda_n$ as in Theorem \ref{thm:neuralAnormality}. Finally, for any $\widetilde\theta, \widetilde\theta'$ on a path between $\theta_*$ and $\theta$, assume that:
$$\nabla_{\widetilde\theta \to \widetilde\theta'-\theta_*}\nabla_{\widetilde\theta \to \theta-\theta_*}\nabla_{\widetilde\theta \to v_*} D_f(\mathbb{P}_{\widetilde\theta}\|\mathbb{P}_{\theta_*})\prec D_f(\mathbb{P}_\theta\|\mathbb{P}_{\theta_*})$$
Then:
\begin{equation}
    \sqrt{n}(F(\theta_n) - F(\theta_*)) \overset{d}{\rightarrow}\mathcal{N}(0, \langle v_*, v_*\rangle)
\end{equation}
\end{proposition}
\begin{remark}\label{rem:parametrictaylor}
In applications, the key difficulty lies in verifying that the third derivative above is bounded by the loss. This condition serves to control the higher order term of the Taylor expansion. Such assumptions are common in the semiparametric literature, e.g. \cite{aichen2003}'s Assumptions 4.5 and 4.6 play the same role. It is easiest to verify in the parametric setting, where 
$$\nabla_{\widetilde\theta \to \widetilde\theta'-\theta_*}\nabla_{\widetilde\theta \to \theta-\theta_*}\nabla_{\widetilde\theta \to v_*} D_f(\mathbb{P}_\theta\|\mathbb{P}_{\theta_*})\asymp \|\theta-\theta_*\|^2_2 \asymp D_f(\mathbb{P}_\theta\|\mathbb{P}_{\theta_*})$$
\end{remark}
Note that $\langle\cdot,\cdot\rangle$ and hence the asymptotics of $\widehat\theta_n$ are independent of $f$, so the $f$-divergences are asymptotically equivalent. An example for a smooth functional $F(\theta)$ that is of particular interest in the semiparametric setting $\theta=(\beta, \alpha)$ is $F(\theta)=\beta'\zeta$, which ``picks out'' a linear combination of the parametric components. This allows us to derive the asymptotic normality of the vector $\sqrt{n}(\widehat\beta-\beta_*)$ in the following Corollary, which makes use of the orthogonal scores assumption that is standard in the semiparametric literature.

\begin{propcorollary}
In addition to the assumptions of Proposition \ref{thm:fdivnormality}, assume the orthogonal scores condition holds: 
$$\mathbb{E} \left[\nabla_{\beta_*\to\beta} \log \D\mathbb{P}_{\beta_*, \alpha_*}(Y)\nabla_{\alpha_*\to\alpha} \log \D\mathbb{P}_{\beta_*, \alpha_*}(Y)\right]=0 \quad \forall \beta,\alpha$$
Then the parametric component $\widehat{\beta}_n$ attains the Cramér-Rao bound:
$$ \sqrt{n}(\widehat\beta_n-\beta_*) \overset{d}{\rightarrow} \mathcal{N}\left(0, I^{-1}\right), \text{ where } I=\mathbb{E}\left[\nabla_{\beta_*}\log \D\mathbb{P}_{\beta_*,\alpha_*}(Y)\cdot\nabla_{\beta_*'}\log \D\mathbb{P}_{\beta_*,\alpha_*}(Y)\right]$$
\end{propcorollary}

\begin{proof}
We simply choose $v_*=(I^{-1}\zeta, \mathbf{0})$, such that $\langle\theta-\theta_*, v_*\rangle=(\beta-\beta_*)'\zeta=F(\theta)-F(\theta_*)$. Since $\langle v_*, v_*\rangle=\zeta' I^{-1}\zeta$, Proposition \ref{thm:fdivnormality} yields $\sqrt{n}(\widehat\beta_n-\beta_*)'\zeta\overset{d}{\to}\mathcal{N}(0,\zeta' I^{-1}\zeta)$. The result then follows via the Cramér-Wold device.
\end{proof}

The $f$-GAN objective therefore attains the efficient asymptotics of maximum likelihood, but does not require explicit knowledge of the model density $\mathbb{P}_\theta$.

\subsection{Generalized Empirical Likelihood} \label{app:GEL}
For the class of Generalized Empirical Likelihood estimators introduced in Section \ref{ex:GEL}, the $\sqrt{n}$-normality and asymptotic efficiency of $\widehat{\theta}_n$ is long established in the parametric case (\cite{imbens1995gel, imbens2002GLE, newey2004generalized}). However, our theoretical framework still allows us to extend the known results to the semiparametric case where $\theta$ may contain unknown functions, which are approximated by a class of neural networks $\Theta_n$ which may grow with $n$. In this case, we can characterize the convergence rate of $\widehat{\theta}_n$ to the identified set  $\Theta_*=\{\theta\in\Theta:\mathbb{E}[m(Y, \theta)]=0\}$, which is unlikely to be singleton given that an infinite-dimensional parameter is hardly pinned down by a finite number of unconditional moment restrictions. We obtain the following result:
\begin{proposition}\label{thm:gelrates}
Let $D_f=\chi^2$ and consider the A-estimator $\widehat{\theta}_n, \widehat{\lambda}_n$ satisfying \ref{assn:outernash},\ref{assn:innernash} with $l(\theta, \lambda , Y)=-f_*(\lambda'm(Y, \theta))$. Let $\Theta_*, \Theta_n$ be as in Theorem \ref{thm:neuralArate} and $\Lambda_*=\Lambda_n=\mathbb{R}^{\operatorname{dim}(m)}$, with $\Bar{r}>\frac{d^*}{p}$. Assume that $m(Y,\theta)-m(Y,\theta')\prec \|\theta-\theta'\|_\infty$ and $|m(Y,\theta)|<\infty$. Then:

$$\mathbb{E}\left[m(Y, \widehat{\theta}_n)\right] = o_\mathbb{P}(n^{-1/(2+\Bar{r})})$$
\end{proposition}
\begin{proof}
We verify the conditions of Theorem \ref{thm:neuralArate}. Assumption A0 holds by assumption, and A1 follows from the Lipschitzness of $m(Y, \cdot)$ and that of $f_*(t)=t+t^2/4$. To verify Assumption 2, note that $l(\theta_*, Y)=0$ and boundedness of $m(Y,\theta)$ imply: 
$$\mathbb{V}[l(\theta,Y)-l(\theta_*, Y)] \prec \mathbb{E}[(m(Y, \theta)'\lambda_*^\theta)^2] \asymp \mathbb{E}[l(\theta, Y)-l(\theta_*, Y)]$$
For the second part of condition A2, simply verify that $\mathbb{E}[l(\theta, Y)-l(\theta_*, Y)]\prec  \|\lambda_*^\theta-\lambda_*^{\theta_*}\|^2_2 \prec \|\theta, \theta_*\|_{\widetilde{\mathcal{X}}}^2 +\mathbb{P}(\Bar{x}\not\in\widetilde{\mathcal{X}})$, which follows by applying the Lipschitzness of $m$ in $\theta$ and the tower-property of $\mathbb{E}$ to $\lambda_*^\theta=-2\mathbb{E}[m(Y,\theta)m(Y, \theta)']^{-1}\mathbb{E}[m(Y,\theta)]$, akin to the proof of \ref{thm:fdivrates}. Assumption A3 can be verified for the Euclidean $\lambda$ via a Taylor expansion, yielding: $\mathbb{V}[l(\theta, \lambda, Y)-l(\theta, \lambda_*^\theta, Y)]\asymp \|\lambda-\lambda_*^\theta\|_2^2\asymp\mathbb{E}[l(\theta, \lambda, Y)-l(\theta, \lambda_*^\theta, Y)]$.
\end{proof}

\subsection{Off-Policy Reinforcement Learning} \label{app:RL}

Next, we will use our theory to the extend the known results about SBEED, the off-policy RL algorithm of \cite{sbeed} introduced in Section \ref{ex:RL}. Theorem \ref{thm:neuralArate} makes it easy to obtain the convergence rates of the corresponding A-estimator:
\begin{proposition}\label{thm:sbeedrates}
Consider the A-estimator $\widehat{\theta}_n, \widehat{\lambda}_n$ satisfying \ref{assn:outernash},\ref{assn:innernash} with $l(\theta, \lambda, Y)$ as in \ref{eq:sbeed}. Assume the observations are iid for simplicity, and that $P_*=P_{\theta_*}$ and $V_*=V_{\theta_*}$, where $\theta_*\in\Theta_*,\lambda_*^\theta\in\Lambda_*$ satisfy A0 in Theorem \ref{thm:neuralArate} with $\mathcal{X}=\Bar{\mathcal{X}}=\mathcal{S}\times\mathcal{A}$. Let $\Theta_*\subset\Theta,\Lambda_*\subset\Lambda$, with $\Theta,\Lambda$ compact under $\|\cdot\|_\infty$ and path-connected. Let $R(\cdot, \cdot), V_\theta(\cdot), P_\theta(\cdot|\cdot)$ be continuous. Let the parametrizations $P_\theta, V_\theta$ satisfy the Lipschitz conditions $\left\|\log P_\theta-\log P_{\theta'}\right\|_\infty\prec \|\theta-\theta'\|_\infty$ and $\left\|V_\theta-V_{\theta'}\right\|_\infty\prec \|\theta-\theta'\|_\infty$. Let the neural network classes $\Theta_n, \Lambda_n$ be constructed as in Theorem \ref{thm:neuralArate}, for any $\underline{r}\geq \frac{d^*}{p} \vee \frac{\Bar{d}^*}{\Bar{p}}$. Then for any $\Bar{r}>\underline{r}$:
$$\mathbb{E}_{s,a}\left[\left(R(s,a) +\beta \mathbb{E}[V_{\widehat{\theta}_n}(s^+)|s, a] - V_{\widehat{\theta}_n}(s)-\log P_{\widehat{\theta}_n}(a|s)\right)^2\right] = o_\mathbb{P}(n^{-2/(2+\Bar{r})})$$
\end{proposition}
\begin{remark}
In contrast to the original work, our result also applies in the case where $\mathcal{A}$ and $\mathcal{S}$ are continuous, and we characterize the optimal rate of growth for the neural network function approximators, which optimally trade off bias and variance. While following almost trivially from the general Theorem \ref{thm:neuralArate}, our result yields significantly faster convergence rates than the $o_\mathbb{P}(\sqrt{n})$ rates obtained by \cite{sbeed}, and our rates further exhibit the reduced curse of dimensionality of neural networks. 
\end{remark}
\begin{remark}
We noticed that SBEED can be viewed as a special case of some of the econometric conditional moment estimators  treated in Example \ref{ex:CMR}, such as \cite{dikkala2020minimax}. We therefore refer the reader to Section \ref{app:CMR} for an application of our asymptotic normality result. Interestingly, neither literature seems to be aware of this connection. \cite{sbeed} cite convex conjugation and the interchangeability principle as the inspiration for their objective, whereas the adversarial conditional moment estimators in econometrics were inspired by \cite{hansen1982gmm}'s Generalized Method of Moments.
\end{remark}

\subsection{A-Estimators for Conditional Moment Restrictions} \label{app:CMR}
We will now apply our theory to examine the asymptotic behavior of the conditional moment estimator of \cite{dikkala2020minimax}, introduced in Section \ref{ex:CMR}. We can apply Theorem \ref{thm:neuralArate} to obtain the rate at which $\widehat{\theta}_n$ converges:
\begin{proposition}\label{thm:cmrrate}
Let $\Theta_n,\Theta_*, \Lambda_n, \Lambda_*$ be as in Theorem \ref{thm:neuralArate}. Let $m(X,\theta)$ be $\|\cdot\|_\infty$-Lipschitz in $\theta$. Let the support of $Y$ be bounded. Then, for any $\Bar{r}>\frac{d^*}{p}\vee\frac{\Bar{d}^*}{\Bar{p}}$, we get:
$$\mathbb{E}\left[\left\|\mathbb{E}[m(X,\widehat{\theta}_n)-m(X,\theta_*)|Z]\right\|^2_2\right]=o_\mathbb{P}(n^{2/(2+\Bar{r})})$$
For the instrumental variable regression setting studied by \cite{dikkala2020minimax}, where $m(X,\theta)=y-\theta(x)$, this implies:
$$\mathbb{E}\left[\left\|\mathbb{E}\left[\widehat{\theta}_n(x)-\theta_*(x)\big\vert Z\right]\right\|^2_2\right]=o_\mathbb{P}(n^{2/(2+\Bar{r})})$$
\end{proposition}
\begin{proof}
Condition A0 is satisfied by assumption, and A1 follows from Lipschitzness of $m(X,\cdot)$ and boundedness. Assumptions A2 and A3 can be verified by using boundedness to establish
\begin{align*}
 \mathbb{V}[l(\theta, Y)-l(\theta_*, Y)] & \prec \|\lambda_*^\theta\|_{\mathcal{L}(\mathcal{Z})^2}^2 \asymp\mathbb{E}[l(\theta,Y)-l(\theta_*, Y)] \\
 \mathbb{V}[l(\theta,\lambda,Y)-l(\theta, \lambda_*^\theta, Y)] & \prec\|\lambda-\lambda_*^\theta\|_{\mathcal{L}(\mathcal{Z})^2}^2\asymp\mathbb{E}[l(\theta,\lambda,Y)-l(\theta, \lambda_*^\theta, Y)]
\end{align*}
\end{proof}
\begin{remark}
Note that just like in the previous Example \ref{ex:GEL}, this result does not require the parameter $\theta_*$ to be identified by the restriction \ref{eq:cmr}. If that is the case however, the above rates can be translated into similar rates in any norm $\|\widehat{\theta}_n-\theta_*\|$ which is dominated by the objective, usually by construction. See \cite{aichen2003} for an example of such a norm in the semi-parameteric setting.
\end{remark}
\begin{remark}
In contrast to \cite{dikkala2020minimax}, our convergence rate result allows for general $m$ and possibly vector-valued, semiparametric $\Theta$ in which unknown functions are approximated by neural networks. Our rates are also exhibit the reduced curse of dimensionality of neural networks.
\end{remark}
Next, we use Theorem \ref{thm:neuralAnormality} to derive the asymptotic variance of the estimator, showing that the estimator is in inefficient in general. For this purpose, it suffices to only consider the simpler parametric setting.
\begin{proposition}\label{thm:cmrnormality}
Consider the parametric case where $\Theta_n=\Theta_*=\Theta$ is Euclidean. In addition to the assumptions of Proposition \ref{thm:cmrrate}, assume that the identification condition \ref{eq:cmr} holds. Let $d(X,\theta):=\nabla_\theta m(X,\theta)$ be bounded and satisfy the Lipschitz condition $|d(X,\theta)-d(X,\theta')|\prec \|\theta-\theta'\|_\infty$. Assume that $\mathbb{E}[l(\theta,Y)]$ is three times differentiable in $\theta$. For all $\theta\in\Theta$, let  $\lambda_*^\theta:=2\mathbb{E}[m(X,\theta)|Z]\in\Lambda_*$ for a $\Lambda_*$ satisfying A0 with $\frac{d^*}{p}<\frac{1}{4}$, let $\theta_*,\lambda_*^\theta$ lie in the respective interiors of $\Theta, \Lambda_*$, and let $\lambda_*^{\prime\theta}[v_*](\cdot):=2v_*'\mathbb{E}[d(X,\theta)|Z=\cdot]\in\Lambda_*$ for any $v_*\in\Theta$. Let $\Lambda_n$ be regularized as in Theorem \ref{thm:neuralAnormality}.  Then:
$$\sqrt{n}(\widehat{\theta}_n-\theta_*) \overset{d}{\to}\mathcal{N}(0, V)$$
where $V=\mathbb{E}\left[\mathbb{E}\left[\nabla_{\theta_*}m(X,\theta_*)'|Z\right]\mathbb{E}\left[m(X,\theta_*)m(X,\theta_*)'|Z\right]\mathbb{E}\left[\nabla_{\theta_*}m(X,\theta_*)|Z\right]\right]^{-1}$.
\end{proposition}
\cite{chamberlain1987asymptotic} derived the efficiency bound for the parametric conditional moment setting, corresponding to the smallest (in a p.s.d. sense) $\sqrt{n}$-asymptotic variance for any unbiased estimator. It is given by the covariance matrix:
$$V_*=\mathbb{E}\left[\mathbb{E}\left[\nabla_{\theta_*}m(X,\theta_*)'|Z\right]\mathbb{E}\left[m(X,\theta_*)m(X,\theta_*)'|Z\right]^{-1}\mathbb{E}\left[\nabla_{\theta_*}m(X,\theta_*)|Z\right]\right]^{-1}$$
Note that $V\neq V_*$ in general, implying that $\widehat{\theta}_n$ is an inefficient estimator. By extension, this also applies to the Reinforcement Learning algorithm of Example \ref{ex:RL}. Comparing the GMM objective of Example \ref{ex:GEL} - which is known to be efficient in the unconditional moment setting - to the population objective of the present example, this may be unsurprising: in contrast to GMM, the population objective of \cite{dikkala2020minimax} corresponds to a regular $\ell^2$ norm, without the inverse covariance weighting which is crucial for asymptotic efficiency in the unconditional case. Generalizing GEL to the conditional moment setting by replacing the constant adversary with a neural network $\Lambda_n$, \cite{metzger2021} therefore proposes the A-estimator given by:
$$\inf_{\theta\in\Theta_n}\sup_{\lambda\in\Lambda_n}\mathbb{E}_n\left[-f_*\left(m(X,\theta)'\lambda(Z)\right)\right]$$
which nests a simplified variant of \cite{variationalMM2020} for $D_f=\chi^2$, and for $D_f=D_{KL}$ can be viewed as alternative to the Kernel approach of \cite{kitamura2004empirical}. \cite{metzger2021} provides a similar information theoretic foundation as the GEL estimator and - building on the theory developed in the present paper - derives the convergence rates and asymptotic efficiency of this estimator, where $\Theta_n$ may contain unknown functions which are modeled as neural networks.

\subsection{Estimating Riesz Representers}
\label{app:riesz}

Finally, we show that Theorem \ref{thm:neuralArate} can be used to quickly derive the convergence rates of \cite{chernozhukov2020adversarial}'s adversarial estimator for Riesz representers, which we introduced in Section \ref{ex:riesz}.
\begin{proposition}\label{thm:rieszrate}
Let $\Theta_n,\Theta_*, \Lambda_n, \Lambda_*$ be as in Theorem \ref{thm:neuralArate}. Let $m(Y,\lambda)=m(Y,\lambda(x))$ be Lipschitz in $\lambda(x)$. Let the support of $Y$ be bounded. Then, for any $\Bar{r}>\frac{d^*}{p}\vee\frac{\Bar{d}^*}{\Bar{p}}$:
$$\|\widehat\theta_n-\theta_*\|_{\mathcal{L}^2(x)}=o_\mathbb{P}(n^{1/(2+\Bar{r})})$$
\end{proposition}
This result clarifies that the Riesz representer of \cite{chernozhukov2020adversarial} can similarly benefit from the adaptivity properties of neural networks, which yield faster rates for our target classes if $d_*<D$. In combination with their Lemma 17, this implies that compared to other non-parametric sieves, neural networks guarantee the asymptotic normality of the orthogonalized estimator $\Tilde\phi_n$ under weaker conditions on smoothness and $D$. Since the normality of $\Tilde\phi_n-\phi_*$ is of primary interest and already follows from \cite{chernozhukov2020adversarial}'s Lemma 17 given our convergence rates, we refrain from deriving it for arbitrary functionals $\widehat\phi_n(g)=\mathbb{E}[\widehat{\theta}_n(x)g(x)]$, although it would be possible to use Theorem \ref{thm:neuralAnormality} to derive $\sqrt{n}(\widehat\phi_n(g)-\phi(g))\overset{d}{\to}\mathcal{N}(0,V_g)$ for some $V_g$ for example.

\section{Conclusion}
We characterize the general class of adversarial estimators {\it (`A-estimators')}, subsuming many estimators independently proposed in the fields of econometrics and machine learning. Our unified framework suggests interesting commonalities between A-estimators: their adversary is always Neyman-orthogonal with respect to the main model, guaranteeing that its estimation errors have no first-order asymptotic impact on the estimated model. Most objectives have versions which asymptotically minimize an $f$-divergence criterion and are asymptotically efficient. Typically, A-estimators adaptively learn how to optimally emphasize the restrictions implied researcher's estimation assumptions, performing particularly well when this set is large. This makes them a promising framework for incorporating machine learning methods into causal inference, where even simple target parameters often satisfy a continuum of restrictions. We characterize the convergence rates of A-estimators, as well as the asymptotic normality of smooth functionals of their parameters. We also provide low-level analogues of these results for semi-parametric models, in which unknown functions are approximated by deep neural networks. Our convergence and normality results also extend the theory of neural network M-estimators, as a special case: building on recent results in approximation theory, our neural network converge rates exhibit a reduced curse of dimensionality for more general losses than previously examined, which hold uniformly over a second parameter space. Our normality result overcomes a problem previously posed by the non-convexity of neural network sieves, showing that a particular regularization, combined with under-smoothing, can be used to satisfy a strong, high-level approximation error condition which the literature left hitherto unverified. 

\bibliography{references}

\appendix
\singlespacing
\section{Definitions}
\label{appendix:definitions}
\begin{definition}[Covering Number]\label{def:coveringnumber}\ \\ 
For some norm $\|\cdot\|$ over some metric space $\Lambda$, the covering number $\mathcal{N}(\delta, \Lambda, \|\cdot\|)$ is defined as the cardinality of the smallest set $C\subset\Lambda$ such that $\sup_{\lambda\in \Lambda}\inf_{c\in C} \|\lambda-c\|\leq\delta$. The quantity $
\log\mathcal{N}(\delta, \Lambda, \|\cdot\|)$ is also called metric entropy.
\end{definition}

\begin{definition}[Deep Neural Networks]\label{def:neuralnets}\ \\
We define the class of deep $\sigma$ networks $f\in \mathcal{F}_{\sigma}(L, W, w, \kappa, B)$ as parametrized functions of the form:
\[f(x)=A^{(L)} \cdot \sigma\left(A^{(L-1)} \cdots \sigma\left(A^{(1)} x+b^{(1)}\right) \cdots+b^{(L-1) }\right)+b^{(L)}\]

where the $A^{(l)}$'s are weight matrices and $b^{(l)}$'s are intercept vectors with real-valued elements, and $\sigma: \mathbb{R}\mapsto \mathbb{R}$ is applied element-wise. For example, the choice $\sigma(x)=\operatorname{ReLU}(x)=\max \{0, x\}$ (rectified linear unit) gives rise to the class of deep ReLU networks, and $\sigma(x)=\tanh{(x)}$ gives rise to the class of $\tanh$ networks. We say the network is $L$ layers deep and call the upper bound $\sup_{l} \operatorname{dim}(b^{(l)})\leq w$ its width. Further, we assume that \[\max_{i, j, l}\left|A^{(l)}_{i j}\right| \leq \kappa,\max_{i, l} |b^{(l)}_i| \leq \kappa, \sum_{l=1}^{L}\left\|A^{(l)}\right\|_{0}+\left\|b^{(l)}\right\|_{0} \leq W, \text { for } i=1, \ldots, L\]

i.e. all elements in the $A^{(l)}$'s and $b^{(l)}$'s are bounded in absolute value by $\kappa$, and there are at most $W$ non-zero parameters in total. Finally, we assume $\|f\|_\infty \leq B<\infty$ for all $f$. If the particular value $B$ is an arbitrary large enough constant, we may suppress the notation and write $\mathcal{F}_{\sigma}(L, W, w, \kappa, B)=\mathcal{F}_{\sigma}(L, W, w, \kappa)$.
\end{definition}

\begin{definition}[Minkowski Dimension]\label{def:minkowski}\ \\
The (upper) Minkowski dimension of a set ${\mathcal{X}} \subset[0,1]^{D}$ is defined as
\[\operatorname{dim}_{M} {\mathcal{X}}:=\inf \left\{d^{*} \geq 0 \mid \limsup _{\varepsilon \downarrow 0} \mathcal{N}(\varepsilon, {\mathcal{X}}, \|\cdot\|_\infty) \varepsilon^{d^{*}}=0\right\}\]
where $\mathcal{N}(\varepsilon, {\mathcal{X}}, \|\cdot\|_\infty)$ is given by Definition \ref{def:coveringnumber}. As shown in \cite{minkowski}, this definition generalizes many other notions of intrinsic dimension, such as the manifold dimension. 
\end{definition}

\begin{definition}[Hölder Space]\label{def:holder}\ \\
For a function $f: \mathbb{R}^{D} \rightarrow \mathbb{R}, \partial_{d} f(x)$ is a partial derivative with respect to a $d$-th component, and $\partial^{\alpha} f:=\partial_{1}^{\alpha_{1}} \cdots \partial_{D}^{\alpha_{D}} f$ using multi-index $\alpha=\left(\alpha_{1}, \ldots, \alpha_{D}\right) .$ For $z \in \mathbb{R}$
$\lfloor z\rfloor$ denotes the largest integer that is less than $z$. Let $p>0$ be a degree of smoothness. For $f:[0,1]^{D} \rightarrow \mathbb{R},$ the Höder norm is defined as
$$
\|f\|_{\mathcal{H}\left( p ,[0,1]^{D}\right)}:=\max _{\alpha:\|\alpha\|_{1}<\lfloor p \rfloor} \sup _{x \in[0,1]^{D}}\left|\partial^{\alpha} f(x)\right|+\max _{\alpha:\|\alpha\|_{1}=\lfloor p \rfloor x, x^{\prime} \in[0,1]^{D}, x \neq x^{\prime}} \frac{\left|\partial^{\alpha} f(x)-\partial^{\alpha} f\left(x^{\prime}\right)\right|}{\left\|x-x^{\prime}\right\|_{\infty}^{ p -\lfloor p \rfloor}}
$$
Then, the Hölder space on $[0,1]^{D}$ is defined as
$$
\mathcal{H}\left( p ,[0,1]^{D}\right)=\left\{f \in C^{\lfloor p \rfloor}\left([0,1]^{D}\right) \mid\|f\|_{\mathcal{H}\left( p ,[0,1]^{D}\right)}<\infty\right\}
$$
Also, $\mathcal{H}\left( p ,[0,1]^{D}, M\right)=\left\{f \in \mathcal{H}\left( p ,[0,1]^{D}\right) \mid\|f\|_{\mathcal{H}\left( p ,[0,1]^{D}\right)} \leq M\right\}$ denotes the $M$-radius closed
ball in $\mathcal{H}\left( p ,[0,1]^{D}\right)$.
\end{definition}

\begin{definition}[(p, C)-smoothness]\label{def:pCsmooth}\ \\
Let $p=q+s$ for some $q \in \mathbb{N}_0$ and $0<s \leq 1$. A function $m: \mathbb{R}^{d} \rightarrow \mathbb{R}$ is called $(p, C)-$smooth, if for every $\alpha=\left(\alpha_{1}, \ldots, \alpha_{d}\right) \in \mathbb{N}_0^{d}$ with
$\sum_{j=1}^{d} \alpha_{j}=q$ the partial derivative $\frac{\partial^{q} m}{\partial x_{1}^{\alpha} \ldots \partial x_{d}^{\alpha_{d}}}$ exists and satisfies
$$
\left|\frac{\partial^{q} m}{\partial x_{1}^{\alpha_{1}} \cdots \partial x_{d}^{\alpha_{d}}}(x)-\frac{\partial^{q} m}{\partial x_{1}^{\alpha_{1}} \cdots \partial x_{d}^{\alpha_{d}}}(z)\right| \leq C \cdot\|x-z\|_2^{s}
$$
for all $x, z \in \mathbb{R}^{d}$.
\end{definition}

\begin{definition}[Generalized Hierarchical Interaction Models]\label{def:GHI}\ \\
Let $C\in \mathbb{R}_{\geq0}$, $D \in \mathbb{N}, d^{*} \in\{1, \ldots, D\}$, $m: \mathbb{R}^{D} \rightarrow \mathbb{R}$ and $p=q+s$ for some $q \in \mathbb{N}_0$ and $0<s \leq 1$. 
\begin{itemize}
    \item[a)] We say that $m$ satisfies a generalized hierarchical interaction model of order $d^{*}$ and level $0$ with bound $C$, if there exist $a_{1}, \ldots, a_{d^{*}} \in \mathbb{R}^{D}$ and some $f: \mathbb{R}^{d^{*}} \rightarrow \mathbb{R}$ such that
$$
m(x)=f\left(a_{1}^{T} x, \ldots, a_{d^{*}}^{T} x\right) \quad \text { for all } x \in \mathbb{R}^{D}
$$
and where $f$ is Lipschitz continuous with constant $C$ and all of its partial derivatives of order less than or equal to $q$ are bounded in absolute value by by $C$.
    \item[b)] We say that $m$ satisfies a generalized hierarchical interaction model of order $d^{*}$ and level $l+1$ with bound $C$ if there exist $K \in \mathbb{N}$, $g_{k}: \mathbb{R}^{d^{*}} \rightarrow \mathbb{R}(k=1, \ldots, K)$ and
$f_{1, k}, \ldots, f_{d^{*}, k}: \mathbb{R}^{D} \rightarrow \mathbb{R}(k=1, \ldots, K)$ such that $f_{1, k}, \ldots, f_{d^{*}, k}(k=1, \ldots, K)$
satisfy a generalized hierarchical interaction model of order $d^{*}$ and level $l$ and
$$
m(x)=\sum_{k=1}^{K} g_{k}\left(f_{1, k}(x), \ldots, f_{d^{*}, k}(x)\right) \quad \text { for all } x \in \mathbb{R}^{D}
$$
where $g_{k}$ are Lipschitz continuous with constant $C$ and all of their partial derivatives of order less than or equal to $q$ are bounded by some constant $C$.
    \item[c)] We say that the generalized hierarchical interaction model defined above is $(p, C)$-smooth, if all functions occurring in its definition are $(p, C)$-smooth, cf. Definition \ref{def:pCsmooth}. 
    \item[d)] We define $\mathcal{G}(p, d^{*}, C, [0,1]^D)$ as the class of all functions $m: [0,1]^D \rightarrow \mathbb{R}$ satisfying a $(p, C)$-smooth generalized hierarchical interaction model of order $d^*$ and level $l$ with bound $C$, where $l\leq C$. Since the particular value of $C$ is not important as long as $C<\infty$, we also write $\mathcal{G}(p, d^*, [0,1]^D)$.
\end{itemize}
\end{definition}

\begin{definition}[Pathwise Derivatives]\label{def:derivatives}\ \\ 
For some $\theta\in\Theta$, $\lambda\in\Lambda$ and some functional $l: \Theta\times\Lambda\mapsto \mathbb{R}^d$, we define the first pathwise derivative in the direction $\theta'\in\Theta$ as $$\nabla_{\theta\to\theta'}l(\theta, \lambda):=\lim_{\tau\to 0}\frac{\partial}{\partial \tau} l(\theta+\tau \theta', \lambda)$$ 
for some real number $\tau\in\mathbb{R}$. Throughout this paper, the usage of $\nabla_{\theta\to\theta'}$ implicitly assumes that the derivative and limit on the RHS exists and is linear in $\theta'$. 
\end{definition}

\section{Supporting Lemmas}
\label{sec:lemmas}
\begin{lemma}[Covering Number of Neural Networks]\label{lem:neuralentropy}\ \\
Consider the class of deep neural networks $f\in \mathcal{F}_{\sigma}(L, W, w, \kappa)$ (Definition \ref{def:neuralnets}), with activation $\sigma$ satisfying $\sigma: |\sigma(x)|\leq x, |\sigma(x)-\sigma(x')|\leq |x-x'| \ \forall x,x'\in\mathbb{R}$ (e.g. ReLU, tanh) and consider the norm $\|f\|_\infty=\sup_{x\in {\mathcal{X}}} |f(x)|$ for some ${\mathcal{X}}\subset [0, 1]^D$ where $D\leq w$. Its $\delta$-covering number (Definition \ref{def:coveringnumber}) can be bounded by:

\[\mathcal{N}\left(\delta, \mathcal{F}_{\sigma}(L, W, w, \kappa),\|\cdot\|_{\infty}\right) \leq\left(\frac{2 L^{2}(w+2)(\kappa w)^{L+1}}{\delta}\right)^{W}\]

\end{lemma}
\begin{proof}
This is Lemma 7 in \cite{chen2020statistical}. While they only state the Lemma for the case of ReLU networks $\sigma(x)=\max(0, x)$, their proof works for any activation $\sigma$ satisfying $|\sigma(x)|\leq x$ and $|\sigma(x)-\sigma(x')|\leq |x-x'|$ for all $x, x'\in\mathbb{R}$. We substituted the bound $B=1$ and renamed some variables.
\end{proof}

\begin{lemma}[Approximation by Deep ReLU Networks on Low Dimensional Data]\label{lem:approxminkowski}\ \\
Consider the Hölder space $\mathcal{H}\equiv\mathcal{H}\left(p,[0,1]^{D}\right)$ (Definition \ref{def:holder}) and some support ${\mathcal{X}}\subset [0,1]^{D}$ with Minkowski dimension (Definition \ref{def:minkowski}) bounded by $\operatorname{dim}_M {\mathcal{X}} \leq d^*\leq D$. For any small enough $\epsilon>0$, the class of deep ReLU networks $\mathcal{F}\equiv\mathcal{F}_{\mathrm{ReLU}}(L, W(\epsilon), w(\epsilon), \kappa(\epsilon))$ (Definition \ref{def:neuralnets}) satisfies: \[\sup_{f_*\in \mathcal{H}} \newinf_{f\in\mathcal{F}} \sup_{x\in {\mathcal{X}}}  |f(x)-f_*(x)| < \epsilon\] as long as $W(\epsilon) \geq c_1 \epsilon^{-d^{*}/ p },\ w(\epsilon) \geq c_2 \epsilon^{-d^{*}/ p },\ \kappa(\epsilon)\geq c_3 \epsilon^{-c_4}$ for any large enough choice of $L$, $c_1$, $c_2$, $c_3$, $c_4>0$. 
\end{lemma}
\begin{proof}
The case $d^*< D$ is covered by Theorem 5 in \cite{minkowski}. While they do not state a bound on the width $w(\epsilon)$, it is easy to see that any network described by Definition \ref{def:neuralnets} with at most $W(\epsilon)$ non-zero parameters can be represented by a network with width bounded by $w(\epsilon)\leq W(\epsilon)$. In the case of $d^*=D$, the Lemma simply states the approximation error for conventional Hölder spaces as established in \cite{yarotsky2017error}.
\end{proof}

\begin{lemma}[Approximation of Generalized Interaction Models by Deep ReLU Networks]\label{lem:approxGHI}\ \\
Consider the function class $\mathcal{G}\equiv\mathcal{G}(p, d^{*}, [0,1]^D)$ (Definition \ref{def:GHI}d) and consider some arbitrary random variable $x\in [0,1]^D$ with probability measure $\mathbb{P}_x$. For any small enough $\epsilon, \eta>0$, the class of deep tanh networks $\mathcal{F}\equiv\mathcal{F}_{\tanh}(L, W(\epsilon), w(\epsilon), \kappa(\epsilon))$ (Definition \ref{def:neuralnets}) satisfies:
\[\sup_{f_*\in \mathcal{G}} \newinf_{f\in\mathcal{F}} \sup_{x\in {\mathcal{X}}} |f(x)-f_*(x)| < \epsilon\] for some subset ${\mathcal{X}}\subset[0,1]^D$ with $\mathbb{P}_x(x\not\in {\mathcal{X}})\leq\eta$ as long as $W(\epsilon) = c_1 \epsilon^{-d^{*}/ p },\ w(\epsilon) = c_2 \epsilon^{-d^{*}/ p },\ \kappa(\epsilon)=c_3 \epsilon^{-c_4}/\eta$ for any large enough choice of $L$, $c_1$, $c_2$, $c_3$, $c_4>0$. 

\end{lemma}
\begin{proof}
This directly follows from Theorem 3 in \cite{bauer2019deep}, however our notation is greatly simplified by the fact that we are not interested in most of their constants, and that we offloaded most of the assumptions into Definition \ref{def:GHI}. What matters is that the network they construct has a depth that bounded by a constant (their equation (6)), and a number of non-zero parameters that is proportional to what they call $(M_n+1)^{d^*}$ in their Theorem 3 (by their equations (7) and (5) and the definition of $M^*$ in their Theorem 3). Since we assumed bounded support (leaving their $a_n$ as a constant), their bound yields an approximation error of $\epsilon=: c M_n^{-p} $ for some $c>0$, such that the number of non-zero parameters can be bounded as $W(\epsilon_n)=O\left((M_n+1)^{d^*}\right)=O(\epsilon^{-d^*/p})$. They bound $\kappa(\epsilon)$ ($\alpha$ in their notation) in terms of $M_n$ and $\eta$ yielding $\kappa(\epsilon)=O(\epsilon^{-c_4}/\eta)$ for some large enough constant $c_4>0$. Finally, their theorem holds only for activation functions $\sigma$ which satisfy a property they call \textit{N-admissible}. While this is technically not satisfied by $\sigma(x)=\tanh{(x)}$, it is easy to verify that this property is satisfied by the activation function $\widetilde{\sigma}(x)=1/2+\tanh{(x)}/2$. Since for any $\Tilde{f}\in \mathcal{F}_{\widetilde{\sigma}} (L, W, w, \kappa)$ there exists some $f\in \mathcal{F}_{\tanh} (L, W, w, 2\kappa+1/2)$ such that $\Tilde{f}=f$, the same approximation bound holds with $\sigma(x)=\tanh{(x)}$. 
\end{proof}

\begin{lemma}[Empirical Process of Donsker Classes]\label{lem:donsker}\ \\
If $Y\in\mathcal{Y}$ is iid and $\{l(\theta, Y): \theta\in\Theta\}$ is $\mathbb{P}$-Donsker for some $l:\Theta\times\mathcal{Y}\mapsto\mathbb{R}$ satisfying $$\lim_{\theta\in\Theta:\|\theta-\theta_*\|\to 0}\mathbb{E}\left[\left(l(\theta, Y)-l(\theta_*, Y)\right)^2\right] = 0,$$ then
$$\sup_{\theta\in\Theta: \|\theta-\theta_*\|\leq \delta_n} (\mathbb{E}-\mathbb{E}_n)[l(\theta, Y)-l(\theta_*, Y)] = o_\mathbb{P}(n^{-1/2})$$
for any $\delta_n=o_\mathbb{P}(1)$.
\end{lemma}
\begin{proof}
This directly follows from Lemma 1 in \cite{chen2003estimation}.
\end{proof}

\begin{lemma}[Empirical Process Rates for A-Estimators]\label{lem:emproc} \ \\ 
Under the assumptions of Theorem \ref{thm:generalrate}, for any function $f(\theta, \lambda, Y)$ satisfying the following conditions:
\begin{itemize}
    \item For any sequence $e_n\geq0$ and all $\theta\in\Theta,\lambda\in\Lambda$: 
    \begin{align}
        \mathbb{V}[f(\theta, \lambda_*^\theta, Y)-f(\theta_*, \lambda_*^{\theta_*}, Y)] & \prec \mathbb{E}[l(\theta, Y)-l(\theta_*, Y) + e_n]^\gamma \label{assn:aloss2}\\
        \mathbb{V}[f(\theta, \lambda, Y)-f(\theta, \lambda_*^\theta, Y)] & \prec \mathbb{E}[l^\theta(\lambda_*^\theta, Y)-l^\theta(\lambda, Y) + e_n]^\gamma\label{assn:mloss2}
    \end{align}
    at least if the right hand sides are smaller than some $C>0$.
    \item For all small $\varepsilon>0$, we have:
    \begin{equation}\label{assn:entropy}
        \log \mathcal{N}(\varepsilon, \{f(\theta, \lambda, \cdot): \theta\in\Theta_n, \lambda\in\Lambda_n\}, \|\cdot\|_\infty) \prec n^{s}(\varepsilon^{-r}-1)/r 
        \end{equation}
\end{itemize}
we obtain the following empirical processes bounds:
\begin{align*}
     \sup_{\substack{\theta\in\widehat{\Theta}_n}} (\mathbb{E}-\mathbb{E}_n)[f(\theta, \pi_n\lambda_*^\theta, Y)-f(\theta_*, \lambda_*^{\theta_*}, Y)] & =O_\mathbb{P}(n^{-\tau(\gamma, s, r, n)} + \epsilon_n+\eta_n+ \Bar\epsilon_n+\Bar\eta_n+ e_n) \\
     \sup_{\substack{\theta\in\widehat\Theta_n \\ \lambda\in\widehat{\Lambda}_n(\theta)}} (\mathbb{E}-\mathbb{E}_n)[f(\theta, \lambda, Y)-f(\theta, \pi_n\lambda_*^\theta, Y)] & =O_\mathbb{P}(n^{-\tau(\gamma, s, r, n)} + \epsilon_n+\eta_n+ e_n)
\end{align*}
where $\widehat{\Lambda}_n(\theta):=\{\lambda\in\Lambda_n:\mathbb{E}[l^\theta(\lambda_*^\theta,Y)-l^\theta(\lambda, Y)]\prec \mathbb{E}[l^\theta(\lambda_*^\theta, Y)-l^\theta(\widehat{\lambda}_n^\theta,Y)]\}$ and $\widehat{\Theta}_n:=\{\theta\in\Theta: \mathbb{E}[l(\theta, Y)-l(\theta_*, Y)]\prec \mathbb{E}[l(\widehat{\theta}_n, Y)-l(\theta_*, Y)]\}$ are shrinking neighborhoods around $\lambda_*^\theta$ and $\theta_*$ containing $\widehat{\lambda}_n^\theta$ and $\widehat{\theta}_n$.

\end{lemma}

\section{Proofs}
\label{appendix:proofs}
\subsection{Theorem \ref{thm:generalrate} and Lemma \ref{lem:emproc}}\label{proof:convergence}
Theorem \ref{thm:generalrate} and Lemma \ref{lem:emproc} are simplified versions of the slightly more general Theorems \ref{thm:Mestimators} and  \ref{thm:Aestimators}, which modify \cite{shen1994}'s M-estimator convergence rate arguments to hold uniformly over another parameter space and accommodate estimators which are finite-sample optimal up to some stochastic remainder. Theorem \ref{thm:Mestimators} is presented in \ref{subsubsec:mrates} and  derives the uniform convergence rates for $\widehat{\lambda}_{n}^\theta$. Theorem \ref{thm:Aestimators} is presented in \ref{subsubsec:arates} and derives the rates for $\widehat{\theta}_n$. In \ref{subsubsec:ratesummary}, we then discuss how Theorem \ref{thm:generalrate} and Lemma \ref{lem:emproc} follow from these results.

\subsubsection{Uniform convergence rate of $\widehat{\lambda}_{n}^\theta$} \label{subsubsec:mrates}
\begin{theorem}[Uniform Convergence Rates of Sieve M-Estimators]\label{thm:Mestimators} Let $\rho^\theta(\cdot,\cdot)$ be a pseudo-distance on $\Lambda$, possibly indexed by $\theta\in\Theta$. For the estimator $\widehat{\lambda}_{n}^\theta$ of \ref{def:Mestimator}, assume:
\begin{itemize}
\item[] CONDITION C1a. For some constants $A_{1}>0$ and $\alpha>0,$ and all small $\varepsilon>0$: $$\inf _{\left\{\rho^\theta\left(\lambda, \lambda_*^\theta\right) \geq \varepsilon, \lambda \in \Lambda,\theta\in\Theta\right\}} \mathbb{E}\left[l^\theta\left(\lambda_*^\theta, Y\right)-l^\theta(\lambda, Y)\right] \geq 2 A_{1} \varepsilon^{2 \alpha} $$
\item[] CONDITION C1b. For some constants $A_{2}>0$ and $\beta>0,$ and all small $\varepsilon>0$: $$\sup _{\left\{\rho^\theta\left(\lambda, \lambda_*^\theta\right) \leq \varepsilon,\lambda \in \Lambda,\theta\in\Theta\right\}} \mathbb{V}\left[l^\theta(\lambda, Y)-l^\theta\left(\lambda_*^\theta, Y\right)\right] \leq A_{2} \varepsilon^{2 \beta} $$
\item[] CONDITION C2. Let $\mathcal{F}_{n}=\left\{l^\theta(\lambda, \cdot)-l^\theta\left(\pi_{n} \lambda_*^\theta, \cdot\right): \lambda \in \Lambda_{n},\theta\in\Theta_n\right\}$.  For some $r_0<\frac{1}{2}$, $A_{3}>0$ and all small $\varepsilon>0$, its entropy (Def. \ref{def:coveringnumber}) is bounded as:
\[\log\mathcal{N}\left(\varepsilon, \mathcal{F}_{n}, \|\cdot\|_\infty\right) \leq A_{3} n^{2 r_0} \varepsilon^{-r}\] where either $r>0$ or $r=0^+$, which is understood to represent $\varepsilon^{-0^+}=\log(1/\varepsilon)$.
\end{itemize}
Let $\epsilon_n := \sup_{\theta\in\Theta_n} \rho^\theta\left(\pi_{n} \lambda_*^\theta, \lambda_*^\theta\right) \vee \left| \mathbb{E}\left[l^\theta(\lambda_*^\theta, Y)-l(\pi_{n} \lambda_*^\theta, Y)\right]\right|^{1/2\alpha}$, then
$$\sup_{\theta\in\Theta_n} \rho^\theta\left(\widehat{\lambda}_{n}^\theta, \lambda_*^\theta\right)=O_{\mathbb{P}}\left(n^{-\tau} + \epsilon_n + \eta_{n}^{1/2\alpha}\right),$$
where $\tau=\tau(\alpha, \beta, r, r_0, n)$ is given by:
$$
\tau=\left\{\begin{array}{ll}
\frac{1-2 r_0}{2 \alpha}-\frac{\log \log n}{2 \alpha \log n}, & \text { if } r=0^{+}, \beta \geq \alpha \\
\frac{1-2 r_0}{4 \alpha-2 \beta}, & \text { if } r=0^{+}, \beta<\alpha \\
\frac{1-2 r_0}{4 \alpha-\min (\alpha, \beta)(2-r)}, & \text { if } 0<r<2 \\
\frac{1-2 r_0}{4 \alpha}-\frac{\log \log n}{2 \alpha \log n}, & \text { if } r=2 \\
\frac{1-2 r_0}{2 \alpha r}, & \text { if } r>2
\end{array}\right.
$$
And for any $f(\theta, \lambda, Y)$ satisfying C1a and C2 when $l^\theta(\lambda, Y)$ is replaced by $f(\theta, \lambda, Y)$, we can bound the empirical process as follows:
\begin{equation}
    \sup_{\substack{\theta\in\Theta_n, \lambda\in\widehat{\Lambda}_n(\theta)}} (\mathbb{E}-\mathbb{E}_n) [f(\theta, \lambda, Y)-f(\theta, \pi_n\lambda_*^\theta, Y)] = O_\mathbb{P}(n^{-\tau} + \epsilon_n + \eta_{n}^{1/2\alpha})
\end{equation}
where $\widehat{\Lambda}_n(\theta)$ is defined as in Lemma \ref{lem:emproc}.
\end{theorem}

\begin{proof}

The theorem generalizes Theorem 1 of \cite{shen1994} such that it holds uniformly over a family of losses indexed by the parameter $\theta\in\Theta_n$, and to allow for the finite-sample optimum to hold approximately up to a possibly stochastic sequence $\eta_n$. Fortunately, the proof can remain almost identical. \cite{shen1994} prove the Theorem by induction, through a chaining argument. They use their Lemma 2 to derive an initial, slow rate which corresponds to the induction start, yielding the assumptions of their Lemma 3 at step $k=2$. Next, their Lemma 3 is repeatedly applied as the induction steps until the rates of Theorem \ref{thm:Mestimators} are obtained. We do not reproduce these algebraic steps, as they are the same as in \cite{shen1994}. Like \cite{shen1994}, we also do not provide the proof for the induction start as it is similar, but simpler than the proof of the induction step, which we present in Lemma \ref{lem:shenstep}.
\end{proof}

\begin{lemma}[Induction Step for Theorem \ref{thm:Mestimators}]\label{lem:shenstep}\ \\
Suppose Conditions C1a, C1b and C2 hold. If at Step $k-1$ we have a rate
$\varepsilon_{n}^{(k-1)}=n^{-\alpha_{k}-1}>\max \left(n^{-\left(1-2 r_0\right) /[\alpha(r+2)]}, \epsilon_n\right)$ so that
\begin{equation*}
\mathbb{P}\left(\sup_{\theta\in\Theta_n}\rho^\theta\left(\widehat{\lambda}_{n}^\theta, \lambda_*^\theta\right) \geq D \varepsilon_{n}^{(k-1)}\right) \leq 5\left[\exp \left(-(1-\varepsilon) \max \left(D^{4 \alpha}, D^{2 \alpha}\right) M_{1} n^{2 r_0}\right)+(k-1) \exp \left(-L n^{\delta_*}\right)\right]
\end{equation*}
where $\delta_*=\min \left(\frac{r+4 r_0}{r+2}, \frac{\beta r\left(1-2 r_0\right)}{4 \alpha}+r_0\right)$ and $L=(1-\varepsilon) \min \left(M_{2} D^{2 \alpha}, M_{3} D^{4 \alpha-\beta(2-r) / 2}\right)$ Then at Step $k,$ we can find an improved rate
\begin{equation*}
\varepsilon_{n}^{(k)}=\max \left(n^{-\alpha_{k}}, n^{-\left(1-2 r_0\right) /[\alpha(r+2)]}, \epsilon_n, \eta_n^{1/2\alpha}\right)
\end{equation*}
where $\alpha_{k}=\left(1-2 r_0\right) /(4 \alpha)+\alpha_{k-1} \beta(2-r) /(4 \alpha),$ so that
\begin{equation*}
\mathbb{P}\left(\sup_{\theta\in\Theta_n}\rho^\theta\left(\widehat{\lambda}_{n}^\theta, \lambda_*^\theta\right) \geq D \varepsilon_{n}^{(k)}\right) \leq 5\left[\exp \left(-(1-\varepsilon) \max \left(D^{4 \alpha}, D^{2 \alpha}\right) M_{1} n^{2 r_0}\right)+k \exp \left(-L {n}^{\delta_*}\right)\right]
\end{equation*}
And for any function $f(\theta, \lambda, Y)$ satisfying C1b and C2 when $l^\theta(\lambda, Y)$ is replaced by $f(\theta, \lambda, Y)$, and $\widehat{\Lambda}_n(\theta)$ is defined as in Lemma \ref{lem:emproc}, the same bound applies to: $$\mathbb{P}\left(\sup_{\substack{\theta\in\Theta_n , \lambda\in\widehat{\Lambda}_n(\theta)}} (\mathbb{E}-\mathbb{E}_n) [f(\theta, \lambda, Y)-f(\theta, \pi_n\lambda_*^\theta, Y)] \geq A_{1} \left(D \varepsilon_{n}^{(k)}\right)^{2 \alpha}\right)$$
\end{lemma}
\begin{proof}
We assume $D>1$ (wlog) and we only prove the case of $4 \alpha \geq \beta(2-r) / 2 .$ Let $B_{n}^{(i)}=\left\{D \varepsilon_{n}^{(i)} \leq \sup_{\theta\in\Theta_n}\rho^\theta\left(\widehat{\lambda}_{n}^\theta, \lambda_*^\theta\right) <D \varepsilon_{n}^{(i-1)}\right\}$ for $i=2, \ldots, k$.
Then
$$
\mathbb{P}\left(\sup_{\theta\in\Theta_n}\rho^\theta\left(\widehat{\lambda}_{n}^\theta, \lambda_*^\theta\right) \geq D \varepsilon_{n}^{(k)}\right) \leq \mathbb{P}\left(\sup_{\theta\in\Theta_n}\rho^\theta\left(\widehat{\lambda}_{n}^\theta, \lambda_*^\theta\right) \geq D \varepsilon_{n}^{(k-1)}\right)+\mathbb{P}\left(B_{n}^{(k)}\right)
$$
To prove the Lemma , we only need to tackle $\mathbb{P}\left(B_{n}^{(k)}\right)$. By Condition C1a,
$$
\begin{array}{l}
\inf _{\left\{\rho^\theta\left(\lambda, \lambda_*^\theta\right) \geq D \varepsilon_{n}^{(k)}, \lambda \in \Lambda_{n},\theta\in\Theta_n\right\}} \mathbb{E}\left[l^\theta\left(\pi_{n} \lambda_*^\theta, Y\right)-l^\theta(\lambda, Y)\right]-\eta_{n} \\
\quad \geq 2 A_{1}\left(D \varepsilon_{n}^{(k)}\right)^{2 \alpha}-\sup_{\theta\in\Theta_n}\mathbb{E}\left[l^\theta\left(\lambda_*^\theta, Y\right)-l^\theta\left(\pi_{n} \lambda_*^\theta, Y\right)\right]-\eta_{n} \geq A_{1}\left(D \varepsilon_{n}^{(k)}\right)^{2 \alpha}
\end{array}
$$
The last inequality requires $A_{1}\left(D \varepsilon_{n}^{(k)}\right)^{2 \alpha}-\sup_{\theta\in\Theta_n}\mathbb{E}\left[l^\theta\left(\lambda_*^\theta, Y\right)-l^\theta\left(\pi_{n} \lambda_*^\theta, Y\right)\right] - \eta_n>0$. This is holds for $A_1>2$ (wlog), which follows from $\varepsilon_{n}^{(k)}\geq \epsilon_n$ and $\varepsilon_{n}^{(k)}\geq \eta_n^{1/2\alpha}$. Thus
\begin{equation*}
\begin{aligned}
\mathbb{P}\left(B_{n}^{(k)}\right) 
& \leq \mathbb{P} \left(\sup_{\left\{D \varepsilon_{n}^{(k)} \leq \rho^\theta\left(\lambda, \lambda_*^\theta\right)<D \varepsilon_{n}^{(k-1)}, \lambda \in \Lambda_{n},\theta\in\Theta_n\right\}} \mathbb{E}_n\left[l^\theta(\lambda, Y)-l^\theta\left(\pi_{n} \lambda_*^\theta, Y\right)\right] \geq-\eta_{n}\right) \\
& \leq \mathbb{P}\left( \sup_{\left\{D \varepsilon_{n}^{(k)} \leq \rho^\theta\left(\lambda, \lambda_*^\theta\right)<D \varepsilon_{n}^{(k-1)}, \lambda \in \Lambda_{n},\theta\in\Theta_n\right\}} n^{1 / 2}(\mathbb{E}_n-\mathbb{E})\left[l^\theta(\lambda, Y)-l^\theta\left(\pi_{n} \lambda_*^\theta, Y\right)\right] \geq A_{1} n^{1 / 2}\left(D \varepsilon_{n}^{(k)}\right)^{2 \alpha}\right)
\end{aligned}
\end{equation*}
Let $v_{k}=\sup_{\left\{D \varepsilon_{n}^{(k)} \leq \rho^\theta\left(\lambda, \lambda_*^\theta\right)<D \varepsilon_{n}^{(k-1)}, \lambda \in \Lambda_{n},\theta\in\Theta_n\right\}} \operatorname{Var}\left(l^\theta\left(\pi_{n} \lambda_*^\theta, Y\right)-l^\theta(\lambda, Y)\right)$. By Condition C1b and $\varepsilon_{n}^{(k-1)}\geq \epsilon_n$ we get $v_{k} \leq 4 A_{2}\left(D \varepsilon_{n}^{(k-1)}\right)^{2 \beta}$. Since  $\varepsilon_{n}^{(k)}$ satisfies
$$
\varepsilon_{n}^{(k)}\geq n^{-\min \left(\left(1-2 r_0\right) /(4 \alpha)+\alpha_{k}-1 \beta(2-r) /(4 \alpha),\left(1-2 r_0\right) /[\alpha(r+2)]\right)}
$$
we know that $n^{1 / 2}\left(D \varepsilon_{n}^{(k)}\right)^{2 \alpha} \geq \max \left(c_{1} n^{-\left(2-r-8 r_0\right) /[2(r+2)]}, c_{2}\left(D \varepsilon_{n}^{k-1}\right)^{2 \beta(2-r) / 4} n^{r_0}\right)$
for some constants $c_{1}>0$ and $c_{2}>0$. We can therefore apply \cite{shen1994}'s Lemma 1 and obtain:
$$
\mathbb{P}\left(B_{n}^{(k)}\right) \leq \exp \left(-\psi_{1}\left(A_{1} n^{1 / 2}\left(D \varepsilon_{n}^{(k)}\right)^{2 \alpha}, v_{k}, \mathcal{F}_{n}\right)\right)
$$
The behavior of $\psi_{1}(\cdot)$ can be analyzed via \cite{shen1994}'s Remark 12.\\
(i) If $\left(D \varepsilon_{n}^{(k)}\right)^{2 \alpha} A_{1}>12\left(D \varepsilon_{n}^{(k-1)}\right)^{2 \beta},$ then
\begin{equation*}
\psi_{1}\left(A_{1} n^{1 / 2}\left(D \varepsilon_{n}^{(k)}\right)^{2 \alpha}, v_{k}, \mathcal{F}_{n}\right) \geq \frac{3 A_{1}}{4} n\left(D \varepsilon_{n}^{(k)}\right)^{2 \alpha} \geq M_{2} D^{2 \alpha} n n^{-2\left(1-2 r_0\right) /(r+2)} \geq M_{2} D^{2 \alpha} n^{\left(r+4 r_0\right) /(r+2)}
\end{equation*}
for some constant $M_{2}>0$.
(ii) If $\left(D \varepsilon_{n}^{(k)}\right)^{2 \alpha} A_{1} \leq 12\left(D \varepsilon_{n}^{(k-1)}\right)^{2 \beta}$, then
\begin{equation*}
\begin{aligned}
\psi_{1} & \left(A_{1} n^{1 / 2}\left(D \varepsilon_{n}^{(k)}\right)^{2 \alpha}, v_{k}, \mathcal{F}_{n}\right)  \geq \frac{\left(A_{1} n^{1 / 2}\left(D \varepsilon_{n}^{(k)}\right)^{2 \alpha}\right)^{2}}{4\left(4 A_{2}\right)\left(D \varepsilon_{n}^{(k-1)}\right)^{2 \beta}} \\
& \geq M_{3} D^{4 \alpha-\beta(2-r) / 2}\left(\varepsilon_{n}^{(k-1)}\right)^{2 \beta(2-r) / 2} \frac{n^{r_0}}{\left(\varepsilon_{n}^{(k-1)}\right)^{2 \beta}} \\
& \geq M_{3} D^{4 \alpha-\beta(2-r) / 2}\left(\varepsilon_{n}^{(1)}\right)^{-\beta r} n^{r_0} \geq M_{3} D^{4 \alpha-\beta(2-r) / 2} n^{\beta r\left(1-2 r_0\right) /(4 \alpha)+r_0}
\end{aligned}
\end{equation*}
for some $M_{3}>0$. Hence,
\begin{equation*}
\mathbb{P}\left(B_{n}^{(k)}\right) \leq\left\{\begin{aligned}
5 \exp \left(-(1-\varepsilon) M_{2} D^{2 \alpha} n^{\left(r+4 r_0\right) /(r+2)}\right)& \text { if }\left(D \varepsilon_{n}^{(k)}\right)^{2 \alpha} A_{1}>12\left(D \varepsilon_{n}^{(k-1)}\right)^{2 \beta} \\
5 \exp \left(-(1-\varepsilon) M_{3} D^{4 \alpha-\beta(2-r) / 2} n^{\beta r\left(1-2 r_0\right) /(4 \alpha)+r_0}\right) &\text { if }\left(D \varepsilon_{n}^{(k)}\right)^{2 \alpha} A_{1} \leq 12\left(D \varepsilon_{n}^{(k-1)}\right)^{2 \beta}
\end{aligned}\right.
\end{equation*}
Take $\delta_*$, $L$ and $\varepsilon_{n}^{(k)}$ as defined in the Lemma, and we obtain $\mathbb{P}\left(B_{n}^{(k)}\right) \leq 5 \exp \left(-L {n}^{\delta_*}\right)$.
This yields the convergence rate of $\widehat{\lambda}_n^\theta$. The statement about arbitrary $f(\theta, \lambda, Y)$ follows by applying analogous arguments (starting at the definition of $v_k$) to the expression $f(\theta, \pi_n\lambda_*^\theta, Y)-f(\theta, \lambda, Y)$ instead of $l^\theta(\pi_n\lambda_*^\theta, Y)-l^\theta(\lambda, Y)$.
\end{proof}

\subsubsection{Convergence of $\widehat{\theta}_{n}$} \label{subsubsec:arates}

\begin{theorem}[Convergence Rate of A-Estimators]\label{thm:Aestimators} Consider the family of  M-estimators $\widehat{\lambda}_n^\theta$ defined in \ref{def:Mestimator} and the A-estimator $\widehat{\theta}_{n}$ defined in \ref{def:Aestimator}. Assume that conditions C1b and C2 are satisfied with $\rho^\theta(\lambda, \lambda'):=\left|\mathbb{E}[l^\theta(\lambda, Y)-l^\theta(\lambda', Y)]\right|$ (hence C1a is automatically satisfied with $\alpha=1/2$). Let $\Bar\rho(\cdot, \cdot)$ be some pseudo-distance on $\Theta$. Assume that the following conditions are satisfied:

\begin{itemize}
\item[] CONDITION C1a' For some constants $\Bar A_{1}>0$ and $\Bar \alpha>0$, and all small $\varepsilon>0$: $$\inf_{\{\Bar\rho\left(\theta, \theta_*\right) \geq \varepsilon, \theta\in\Theta_n\}} \mathbb{E}\left[l\left(\theta, Y\right)-l(\theta_*, Y)\right] \geq 2 \Bar A_{1} \varepsilon^{2 \Bar \alpha} $$
\item[] CONDITION C1b'. For some constants $\Bar A_{2}>0$ and $\Bar \beta>0,$ and all small $\varepsilon>0$: $$\sup _{\left\{\Bar\rho\left(\theta, \theta_*\right) \leq \varepsilon,\theta\in\Theta_n\right\}} \mathbb{V}\left[l(\theta, Y)-l\left(\theta_*, Y\right)\right] \leq  \Bar A_{2} \varepsilon^{2\Bar  \beta} $$
\item[] CONDITION C2'. Let $\Bar{\mathcal{F}}_{n}=\left\{l(\theta, \pi_{n} \lambda_*^\theta, \cdot)-l\left(\pi_n \theta_*, \pi_{n} \lambda_*^{\pi_n \theta_*}, \cdot\right): \theta\in\Theta_n\right\}$. For some $\Bar r_0<\frac{1}{2}$, $\Bar A_{3}>0$ and all small $\varepsilon>0$, its entropy (Def. \ref{def:coveringnumber}) is bounded as:
\[\log\mathcal{N}\left(\varepsilon,\Bar{\mathcal{F}}_{n}, \|\cdot\|_\infty\right) \leq\Bar  A_{3} n^{2 \Bar r_0} \varepsilon^{-\Bar r}\] where either $\Bar r>0$ or $\Bar r=0^+$, which represents $\varepsilon^{-0^+}=\log(1/\varepsilon)$.
\end{itemize}
Let $\Bar \epsilon_n :=\rho\left(\pi_{n} \theta_*, \theta_*\right) \vee \left|\mathbb{E}l(\pi_{n} \theta_*,Y) -l(\theta_*, Y)\right|^{1/2\Bar\alpha}$ be the approximation error of $\Theta_n$. Then:
$$\Bar\rho\left(\widehat{\theta}_{n}, \theta_*\right)=O_{\mathbb{P}}\left(n^{-\Bar \tau} + \Bar \epsilon_n + (\Bar \eta_n + n^{-\tau} + \epsilon_n + \eta_{n})^{1/2\Bar\alpha}\right)$$
Where $\tau=\tau(1/2, \beta, r, r_0, n)$ and $\epsilon_n$ are defined as in Thm \ref{thm:Mestimators} and $\Bar\tau=\tau(\Bar\alpha, \Bar\beta, \Bar r, \Bar r_0, n)$. Also, for every $f(\theta,\lambda, Y)$ satisfying  C2 and C3 when $l(\theta, \lambda, Y)$ is replaced by $f(\theta, \lambda, Y)$ (recall $l(\theta, Y)=l(\theta, \lambda_*^\theta, Y)$), we can bound the empirical process:
$$
\sup_{\theta\in\widehat{\Theta}_n} (\mathbb{E}-\mathbb{E}_n)[f(\theta, \pi_n\lambda_*^{\theta}, Y)-f(\theta_*, \lambda_*^{\theta_*}, Y)] = O_{\mathbb{P}}\left(n^{-\Bar \tau} + \Bar \epsilon_n + (\Bar \eta_n + n^{-\tau} + \epsilon_n + \eta_{n})^{1/2\Bar\alpha}\right)
$$
\end{theorem}

\begin{proof}
The proof is similar to that of Theorem \ref{thm:Mestimators}. Again, we will only prove the induction step via Lemma \ref{lem:shenstep2} in the Online Appendix, as the remaining arguments are analogous to the proof of the previous Theorem or that of Theorem 1 in \cite{shen1994}. The proof of Lemma \ref{lem:shenstep2} largely mirrors that of Lemma \ref{lem:shenstep}, but uses the results of Theorem \ref{thm:Mestimators} to control the convergence of the adversary. The main additional complexity lies in properly switching back and forth between the sieve spaces and the target function spaces, when bounding the empirical process terms and variances respectively.
\end{proof}

\subsubsection{Proofs of Theorem \ref{thm:generalrate} and Lemma \ref{lem:emproc}} \label{subsubsec:ratesummary}
\begin{proof}
To see that Theorem \ref{thm:generalrate} and Lemma \ref{lem:emproc} follow from the previous results, simply choose $\rho^\theta(\lambda, \lambda')=|\mathbb{E}[l^\theta(\lambda, Y)-l^\theta(\lambda', Y)]|$ and $\Bar\rho(\theta, \theta')=|\mathbb{E}[l(\theta, Y)-l(\theta', Y)]|$, such that Conditions C1a and C1a' are automatically satisfied with $\alpha=1/2$. Further, substitute $\gamma=2\beta \vee 2\Bar\beta$ such that Conditions C1b and C1b' hold by assumptions \ref{assn:mloss} and \ref{assn:aloss} for all $\varepsilon<1\wedge C$. Conditions C2 and C2' directly follow from \ref{eq:covassn}, substituting $s=2r_0=2\Bar r_0$ and fixing $r=\Bar r$. This yields Theorem \ref{thm:generalrate}, and Lemma \ref{lem:emproc} with $e_n=0$.\\

For a proof of Lemma \ref{lem:emproc} with $e_n\neq 0$, note that Condition C1b in Theorem \ref{thm:Mestimators} is only needed to verify $v_k\leq 4A_2\left(D \varepsilon_n^{(k-1)}\right)^{2\beta}$ in the proof of Lemma \ref{lem:shenstep}. Hence we can re-define $\epsilon_n \gets \epsilon_n + e_n$ such that the definition of $\varepsilon_n^{(k-1)}\geq \epsilon_n$ automatically ensures $v_k\prec \left(D \varepsilon_n^{(k-1)}\right)^{2\beta}$. This change in constants does not affect the result. Analogous arguments can be applied to Lemma \ref{lem:shenstep2} and thus Theorem \ref{thm:Aestimators}.
\end{proof}

\subsection{Theorem \ref{thm:normality}}
\begin{proof}
The approximate Nash conditions \ref{assn:outernash} and \ref{assn:innernash} imply
\begin{align}
    \begin{split}
        O_\mathbb{P}(e_n^2) & \geq \mathbb{E}_n l(\widehat{\theta}_n, \pi_n\Bar{\lambda}_n^{\widehat{\theta}_n}(\widehat{\lambda}_n), Y) - l(\pi_n \Bar{\theta}(\widehat{\theta}_n), \widehat{\lambda}_n, Y) \\
        & = \mathbb{E}_n l'(\widehat{\theta}_n, \widehat{\lambda}_n, Y)[\widehat{\theta}_n-\pi_n\Bar{\theta}_n(\widehat{\theta}_n), \pi_n\Bar{\lambda}_n^{\widehat{\theta}_n}(\widehat{\lambda}_n)-\widehat{\lambda}_n] + O_\mathbb{P}(e_n^2) \\
        & = \mathbb{E}_n l'(\widehat{\theta}_n, \widehat{\lambda}_n, Y)[e_n v, e_n \lambda_*^{\prime\widehat{\theta}_n}[v]] + O_\mathbb{P}(e_n^2)
    \end{split}
\end{align}
The second line uses Taylor's theorem. The third line substitutes the definitions of $\Bar\theta_n,\Bar\lambda_n^{\widehat\theta_n}$ and applies Condition N3. Since the signs of $v, \lambda_*^{\prime\widehat{\theta}_n}[v]$ are arbitrary, we may replace the inequality with an equality, which yields $O_\mathbb{P}(e_n) = \mathbb{E}_n l'(\widehat{\theta}_n, \widehat{\lambda}_n, Y)[v,\lambda_*^{\prime\widehat{\theta}_n}[v]]$. Adding and subtracting a few terms, we get:
\begin{align}
\begin{split}
    \mathbb{E}_n l'(\theta_*, Y)[v] = & (\mathbb{E}_n-\mathbb{E}) [l'(\widehat{\theta}_n,\widehat{\lambda}_n, Y)[v,\lambda_*^{\prime\widehat{\theta}_n}[v]]-l'(\theta_*, Y)[v]] \\
    & + \mathbb{E} [l'(\widehat{\theta}_n,\widehat{\lambda}_n, Y)[v,\lambda_*^{\prime\widehat{\theta}_n}[v]]-l'(\theta_*, Y)[v]] + O_\mathbb{P}(e_n) \\
    = & \langle \widehat{\theta}_n-\theta_*, v\rangle + O_\mathbb{P}(e_n)
\end{split}
\end{align}
Where the last line uses Conditions N1 and N2. Substituting $v=v_*$, we get:
\[\sqrt{n} \left( F(\widehat{\theta}_n)-F(\theta_*)\right) = \sqrt{n}\langle \widehat{\theta}_n-\theta_*, v_*\rangle +o_\mathbb{P}(1) = \sqrt{n}\mathbb{E}_n [l'(\theta_*, Y)[v]]+o_\mathbb{P}(1) \overset{d}{\longrightarrow} \mathcal{N}(0, V)\]
with $V=\mathbb{V}\left(l'(\theta_*, Y)[v]\right)$ by the standard central limit theorem.
\end{proof}

\subsection{Theorem \ref{thm:neuralAnormality}}
\begin{proof}[Proof of Theorem \ref{thm:neuralAnormality}]
Note that the regularization does not interfere with Theorem \ref{thm:neuralArate}: the approximation power relative to $\Theta_*, \Lambda_*$ is not reduced as we remove only elements form the sieves $\Theta_n, \Lambda_n$ that are far away from $\Theta_*, \Lambda_*$, and the regularization is sufficiently slow to guarantee that the sieves are nonempty for small enough $\epsilon>0$. Hence Theorem \ref{thm:neuralArate} holds with $\bar{r}=2$, yielding rates $o_\mathbb{P}(n^{-2/(2+\bar{r})})=o_\mathbb{P}(n^{-1/2})$. Also note that the assumption $\frac{d^*}{p}\vee\frac{\Bar{d}^*}{\Bar{p}}<1/4$ along with the lower bound $ \underline{r}>2/3$ ensures $\sup_{\theta\in\Theta_*}\|\theta-\pi_n\theta\|_{\bar{{\mathcal{X}}}}=o(n^{-1})$ and $\sup_{\lambda\in\Lambda_*}\|\lambda-\pi_n\lambda\|_{{\mathcal{X}}}=o(n^{-1})$. We first verify condition N1, decomposing it into two parts by adding and subtracting $l'(\theta,\lambda_*^{\prime\theta}, Y)[v_*,\lambda_*^{\prime\theta}[v_*]]=l'(\theta, Y)[v_*]$. First, we show
$$
\sup_{\theta\in\widehat{\Theta}_n,\lambda\in\widehat{\Lambda}_n(\theta)} (\mathbb{E}_n-\mathbb{E}) l'(\theta, Y)[v_*] - l'(\theta_*, Y)[v_*]  = o_\mathbb{P}(n^{-1/2})
$$
If A7ii) holds with $\mathbb{V}[l'(\theta_*, Y)[v] - l'(\theta, Y)[v]] \prec \mathbb{E}[l(\theta, Y)-l(\theta_*, Y)]$, then this can be established with Lemma \ref{lem:emproc} for $\gamma=1$, using the Lipschitz condition A4 and analogous arguments to those in the proof of Theorem \ref{thm:neuralArate}. Lemma \ref{lem:emproc} then yields the same $o_\mathbb{P}(n^{-1/2})$ rates as Theorem \ref{thm:neuralArate}. If A7ii) instead asserts that $\Theta_*$ is $\mathbb{P}$-Donsker, the same result follows from Lemma \ref{lem:donsker}, which can be applied because the Lipschitz continuity A4 implies the L2 continuity required by the Lemma. This implies condition N1, together with: 
$$
\sup_{\theta\in\widehat{\Theta}_n,\lambda\in\widehat{\Lambda}_n(\theta)} (\mathbb{E}_n-\mathbb{E}) l'(\theta,\lambda, Y)[v_*,\lambda_*^{\prime\theta}[v_*]]-l'(\theta,\lambda_*^{\prime\theta}, Y)[v_*,\lambda_*^{\prime\theta}[v_*]] = o_\mathbb{P}(n^{-1/2})
$$
which can be established via analogous arguments and A7i). We proceed to verify condition N2. Using a similar decomposition, we note that 
$$
\sup_{\theta\in\widehat{\Theta}_n,\lambda\in\widehat{\Lambda}_n(\theta)} \mathbb{E} l'(\theta,\lambda, Y)[v_*,\lambda_*^{\prime\theta}[v_*]]- l'(\theta,\lambda_*^{\prime\theta}, Y)[v_*,\lambda_*^{\prime\theta}[v_*]]  =  O_\mathbb{P}(e_n) 
$$
which follows from A6i) and the $o_\mathbb{P}(n^{-1/2})$ rates of Theorem \ref{thm:neuralArate}. Similarly, A6ii) implies $\sup_{\theta\in\widehat{\Theta}_n,\lambda\in\widehat{\Lambda}_n(\theta)} \mathbb{E} l'(\theta,Y)[v_*]-l'(\theta_*, Y)[v_*] - \langle \theta-\theta_*, v_* \rangle =  O_\mathbb{P}(e_n)$ hence condition N2 holds. Finally, we verify condition N3. Define $\pi_*\theta:=\arg\inf_{\theta'\in\Theta_*}\|\theta'-\theta\|_\infty$ as the projection onto $\Theta_*$. Similarly, $\pi_*\lambda:=\arg\inf_{\lambda'\in\Lambda_*}\|\lambda'-\lambda\|_\infty$. Due to the reguarlization, we know $\|\theta-\pi_*\theta\|_\infty=o(n^{-1})$. Therefore $\|\Bar{\theta}_n(\theta)-(\pi_*\theta -e_nv_*)\|_\infty=o(n^{-1})$. By convexity of $\Theta_*$, we have $(\pi_*\theta -e_nv_*)\in\Theta_*$ for $n$ large enough. Given that $\sup_{\theta'\in\Theta_*}\|\theta'-\pi_n\theta'\|_\infty=o(n^{-1})$ due to $\frac{d^*}{p}\vee\frac{\Bar{d}^*}{\Bar{p}}<1/4$ and our choice of $\underline{r}$, we get $\|(\pi_*\theta -e_nv_*)-\theta_n(\pi_*\theta -e_nv_*)\|_\infty=o(n^{-1})$. Taken together, these statements imply $\|\Bar{\theta}_n(\theta)-\pi_n\Bar{\theta}_n(\theta)\|_\infty=o(n^{-1})$. Analogous arguments yield $\|\Bar{\lambda}_n^\theta(\lambda)-\pi_n\Bar{\lambda}_n^\theta(\lambda)\|_\infty=o(n^{-1})$. Hence N3 holds.
\end{proof}

\section{Online Appendix}\label{appendix:online}

\subsection{Proof of Theorem \ref{thm:Aestimators}}

The induction step for the proof of Theorem \ref{thm:Aestimators} is given by the following Lemma.
\begin{lemma}[Induction Step for Theorem \ref{thm:Aestimators}]\label{lem:shenstep2}\ \\
Suppose the Conditions of Theorem \ref{thm:Aestimators} hold. If at Step $k-1$ we have a rate
$$
\varepsilon_{n}^{(k-1)}=n^{-{\Bar\alpha}_{k}-1}>\max \left(n^{-\left(1-2 {\Bar r_0}\right) /[{\Bar\alpha}(\Bar r+2)]}, {\Bar\epsilon}_n, \epsilon_n\right)
$$
so that
\begin{equation*}
\mathbb{P}\left(\Bar\rho\left(\widehat{\theta}_{n}, \theta_*\right) \geq D \varepsilon_{n}^{(k-1)}\right) \leq 5\left[\exp \left(-(1-\varepsilon) \max \left(D^{4 {\Bar\alpha}}, D^{2 {\Bar\alpha}}\right) M_{1} n^{2 {\Bar r_0}}\right)+(k-1) \exp \left(-L n^{\delta_*}\right)\right]
\end{equation*}
where $\delta_*=\min \left(\frac{r+4 {\Bar r_0}}{r+2}, \frac{{\Bar\beta} \Bar r\left(1-2 {\Bar r_0}\right)}{4 {\Bar\alpha}}+{\Bar r_0}\right)$ and $L=(1-\varepsilon) \min \left(M_{2} D^{2 {\Bar\alpha}}, M_{3} D^{4 {\Bar\alpha}-{\Bar\beta}(2-\Bar r) / 2}\right)$, then at Step $k$, we can find an improved rate
$$
\varepsilon_{n}^{(k)}=\max \left(n^{-{\Bar\alpha}_{k}}, n^{-\left(1-2 {\Bar r_0}\right) /[{\Bar\alpha}(\Bar r+2)]}, {\Bar\epsilon}_n, ({\Bar\eta}_n+r_n)^{1/2{\Bar\alpha}}\right)
$$
where ${\Bar\alpha}_{k}=\left(1-2 {\Bar r_0}\right) /(4 {\Bar\alpha})+{\Bar\alpha}_{k-1} {\Bar\beta}(2-\Bar r) /(4 {\Bar\alpha}),$ so that
\begin{equation*}
\mathbb{P}\left(\Bar\rho\left(\widehat{\theta}_{n}, \theta_*\right) \geq D \varepsilon_{n}^{(k)}\right) \leq 5\left[\exp \left(-(1-\varepsilon) \max \left(D^{4 {\Bar\alpha}}, D^{2 {\Bar\alpha}}\right) M_{1} n^{2 {\Bar r_0}}\right)+k \exp \left(-L {n}^{\delta_*}\right)\right]
\end{equation*}
Furthermore, for every $f(\theta,\lambda, Y)$ satisfying Conditions C1b and C2 when $l(\theta, \lambda, Y)$ is replaced by $f(\theta, \lambda, Y)$ (recall $l(\theta, Y)=l(\theta, \lambda_*^\theta, Y)$), the same bound applies to:
$$
\mathbb{P}\left(\sup_{\theta\in\widehat{\Theta}_n} (\mathbb{E}-\mathbb{E}_n)[f(\theta, \pi_n\lambda_*^{\theta}, Y)-f(\theta_*,\lambda_*^{\theta_*}, Y)] \geq D \varepsilon_{n}^{(k)}\right)
$$
\end{lemma}
\begin{proof}
As in the proof of Lemma \ref{lem:shenstep} we assume $D>1$ (wlog) and we only prove the case of $4 {\Bar\alpha} \geq {\Bar\beta}(2-\Bar r) / 2$. Let $B_{n}^{(i)}=\left\{D \varepsilon_{n}^{(i)} \leq \Bar\rho\left(\widehat{\theta}_{n}, \theta_*\right) <D \varepsilon_{n}^{(i-1)}\right\}$ for $i=2, \ldots, k$. As before, we only need to bound $\mathbb{P}\left(B_{n}^{(k)}\right)$. To this end, it will be useful to define
$$r_n := \sup_{\theta\in\Theta_n}\mathbb{E}[l(\theta,\lambda_*^\theta, Y)-l(\theta, \widehat{\lambda}_n^\theta, Y)] \vee \sup_{\theta\in\Theta_n}(\mathbb{E}-\mathbb{E}_n)[l(\theta, \widehat{\lambda}_n^\theta, Y)-l(\theta,\pi_n\lambda_*^\theta, Y)]$$
such that Theorem \ref{thm:Mestimators} implies $r_n = O_\mathbb{P}(n^\tau + \epsilon_n + \eta_n)$, 
which also implies, by definition of $r_n$ and $\epsilon_n$: $\sup_{\theta\in\Theta_n}\left|\mathbb{E}_n [l(\theta, \widehat{\lambda}_n^\theta, Y)-l(\theta,\pi_n\lambda_*^\theta, Y)]\right|\leq r_n + \epsilon_n\leq 2 r_n$. Together with \ref{def:Aestimator} this yields:
\begin{equation}
\label{eq:empiricalAloss}
    \mathbb{E}_n [l(\widehat{\theta}_n, \pi_n\lambda_*^{\widehat{\theta}_n}, Y)-l(\theta,\pi_n\lambda_*^\theta, Y)] \leq 2 r_n + \Bar\eta_n \quad \forall\theta\in\Theta_n
\end{equation}
By Condition C1a', we can therefore bound:
\begin{align}
\begin{split}
\inf_{{\theta\in\Theta_n: \Bar\rho\left(\theta, \theta_*\right) \geq D \varepsilon_{n}^{(k)}}} & \mathbb{E}\left[l(\theta, \pi_{n} \lambda_*^\theta, Y)-l(\pi_n \theta_*, \pi_{n} \lambda_*^{\pi_n \theta_*}, Y)\right]-{\Bar\eta}_{n} - 2 r_n\\
 = \inf_{{\theta\in\Theta_n: \Bar\rho\left(\theta, \theta_*\right) \geq D \varepsilon_{n}^{(k)}}} & \mathbb{E}[l(\theta, Y)-l(\theta_*, Y)] + \mathbb{E}[l(\theta_*, Y)-l(\pi_n\theta_*, Y)] -{\Bar\eta}_{n} - 2 r_n\\
& + \mathbb{E}[l(\theta, \pi_{n} \lambda_*^\theta, Y)-l\left(\theta, Y\right)] + \mathbb{E}[l(\pi_n\theta_*, Y)-l(\pi_n \theta_*, \pi_{n} \lambda_*^{\pi_n \theta_*}, Y)]\\
 \geq & 2 {\Bar A}_1\left(D \varepsilon_{n}^{(k)}\right)^{2 {\Bar\alpha}} - \Bar\epsilon_n^{2\Bar\alpha}-{\Bar\eta}_{n} - r_n - \epsilon_n - \epsilon_n \geq  \Bar A_{1}\left(D \varepsilon_{n}^{(k)}\right)^{2 {\Bar\alpha}}
\end{split}
\end{align}
Where the last line used C1a', the definition of the approximation errors, assumed large enough $\Bar A_1$ (wlog) and used various lower-bounds implied by the definition of $\varepsilon_{n}^{(k)}$. Together with \ref{eq:empiricalAloss}, this yields:
\begin{equation*}
\begin{aligned}
\mathbb{P}\left(B_{n}^{(k)}\right) 
& \leq \mathbb{P} \left(\sup_{\left\{D \varepsilon_{n}^{(k)} \leq \Bar\rho\left(\theta, \theta_*\right)<D \varepsilon_{n}^{(k-1)},\theta\in\Theta_n\right\}} \mathbb{E}_n\left[l(\pi_n \theta_*, \pi_{n} \lambda_*^{\pi_n \theta_*}, Y)-l\left(\theta, \pi_{n} \lambda_*^\theta, Y\right)\right] \geq-{\Bar\eta}_{n}- 2 r_n\right)  \\
& \leq \mathbb{P}\left( \sup_{\substack{\theta\in\Theta_n \\ D \varepsilon_{n}^{(k)} \leq \Bar\rho\left(\theta, \theta_*\right)<D \varepsilon_{n}^{(k-1)}}} n^{1 / 2}(\mathbb{E}_n-\mathbb{E})\left[l(\pi_n \theta_*, \pi_{n} \lambda_*^{\pi_n \theta_*}, Y)-l\left(\theta, \pi_{n} \lambda_*^\theta, Y\right)\right] \geq A_{1} n^{1 / 2}\left(D \varepsilon_{n}^{(k)}\right)^{2 {\Bar\alpha}}\right)
\end{aligned}
\end{equation*}
Let $v_{k}=\sup_{\left\{D \varepsilon_{n}^{(k)} \leq \Bar\rho\left(\theta, \theta_*\right)<D \varepsilon_{n}^{(k-1)}, \theta\in\Theta_n\right\}}\mathbb{V}\left[l(\pi_n \theta_*, \pi_{n} \lambda_*^{\pi_n \theta_*}, Y)-l\left(\theta, \pi_{n} \lambda_*^\theta, Y\right)\right]$. To bound $v_k$, we add and subtract terms and apply the Cauchy-Schwartz inequality:
\begin{equation*}
\begin{aligned}
    %\begin{split}
        \mathbb{V} & [l(\pi_n \theta_*, \pi_{n} \lambda_*^{\pi_n \theta_*}, Y)-l(\theta, \pi_{n} \lambda_*^\theta, Y)] \\
        & \leq 3  \mathbb{V}[l(\pi_n \theta_*, \lambda_*^{\pi_n \theta_*}, Y)-l(\theta, \lambda_*^\theta, Y)] + 3 \mathbb{V}[l(\pi_n \theta_*, \pi_{n} \lambda_*^{\pi_n \theta_*}, Y)-l(\theta, \pi_{n} \lambda_*^\theta, Y)]\\
        & \prec \mathbb{V}[l(\theta, Y) - l(\theta_*, Y)]+\mathbb{V}[l(\pi_n \theta_*, Y) - l(\theta_*, Y)] + \mathbb{V}[l^{\pi_n \theta_*}(\pi_{n} \lambda_*^{\pi_n \theta_*}, Y)-l^{\pi_n \theta_*}(\lambda_*^{\pi_n \theta_*}, Y)] \\
        & \quad + \mathbb{V}[l^\theta(\lambda_*^\theta, Y)-l^\theta(\pi_{n} \lambda_*^\theta, Y)]
    %\end{split}
\end{aligned}
\end{equation*}
By Conditions C1b and C1b', and since $\varepsilon_{n}^{(k-1)}\geq {\Bar\epsilon}_n$, we obtain $v_{k} \leq 4 \Bar A_{2}\left(D \varepsilon_{n}^{(k-1)}\right)^{2 {\Bar\beta}}$, assuming $\Bar A_2$ is large enough (wlog). The remaining arguments are unchanged from the proof of Lemma \ref{lem:shenstep}, which eventually yields: $\mathbb{P}\left(B_{n}^{(k)}\right) \leq 5 \exp \left(-L {n}^{\delta_*}\right)$. This completes the proof for the convergence rate of $\widehat{\theta}_n$. To prove the statement about arbitrary $f(\theta, \lambda, Y)$ satisfying C1b and C2, simply repeat the arguments of the previous proof (starting at the definition of $v_k$) with $l(\theta, \lambda, Y)$ replaced by $f(\theta, \lambda, Y)$. 
\end{proof}

\subsection{Proof of Proposition \ref{thm:fdivrates}}
\begin{proof}
The first order conditions for $\lambda$ in \ref{eq:fdiv} yield the optimal population adversary $\lambda_*^\theta(y)$:
\begin{equation}
    \lambda_*^\theta(y) \D\mathbb{P}_\theta(y) - f_*'(\lambda_*^\theta(y)) \D\mathbb{P}(y) \overset{!}{=} 0 \iff \lambda_*^\theta(y) = f'\left(\frac{\D\mathbb{P}_\theta(y)}{\D\mathbb{P}(y)}\right)
\end{equation}

We verify the conditions of Theorem \ref{thm:neuralArate}. The Lipschitz condition A1 can be verified by writing $l(\theta,\lambda, Y)=\int \lambda(y) \frac{\D\mathbb{P}_\theta}{\D\mathbb{P}}(y) \D\mathbb{P}(y) - f_*(\lambda(Y))$ and using the Lipschitzness of $ \frac{\D\mathbb{P}_\theta}{\D\mathbb{P}}$ in $\theta$ and that of $f_*$ (which follows from boundedness of $\Lambda$ and differentiability of $f$). Towards A2, apply a 2nd order Taylor expansion (with mean value reminder) of $D_f(\mathbb{P}_\theta\|\mathbb{P})$ at $\frac{\D\mathbb{P}_\theta}{\D\mathbb{P}}=\frac{\D\mathbb{P}_{\theta_*}}{\D\mathbb{P}}$ in direction of $\frac{\D\mathbb{P}_\theta}{\D\mathbb{P}}$, which yields for some $\widetilde{\theta}\in\Theta$ on a path from $\theta_*$ to $\theta$:
\begin{equation*}
D_f(\mathbb{P}_\theta\|\mathbb{P}) = \int f\left(\frac{\D\mathbb{P}_\theta}{\D\mathbb{P}}\right) \D\mathbb{P} = \frac{1}{2}\int \left(\frac{\D\mathbb{P}_\theta}{\D\mathbb{P}}-\frac{\D\mathbb{P}_{\theta_*}}{\D\mathbb{P}}\right)^2 f''\left(\frac{\D\mathbb{P}_{\widetilde{\theta}}}{\D\mathbb{P}}\right)\D\mathbb{P}\asymp \left\|\frac{\D\mathbb{P}_\theta}{\D\mathbb{P}}-\frac{\D\mathbb{P}_{\theta_*}}{\D\mathbb{P}}\right\|_{L^2(\mathcal{Y})}^2
\end{equation*}
where the last step follows from strict positivity and boundedness of $f''\left(\frac{\D\mathbb{P}_\theta}{\D\mathbb{P}}\right)$ wp1. Further note that 
$$\mathbb{V}[l(\theta, Y)-l(\theta_*, Y)]=\mathbb{V}\left[f_*\left(\frac{\D\mathbb{P}_\theta}{\D\mathbb{P}}(Y)\right)-f_*\left(\frac{\D\mathbb{P}_{\theta_*}}{\D\mathbb{P}}(Y)\right)\right]\prec \left\|\frac{\D\mathbb{P}_\theta}{\D\mathbb{P}}-\frac{\D\mathbb{P}_{\theta_*}}{\D\mathbb{P}}\right\|_{L^2(\mathcal{Y})}^2$$
due to Lipschitzness of $f_*$. Also note that Lipschitzness of $\frac{\D\mathbb{P}_\theta}{\D\mathbb{P}}$ in $\theta$ implies
$$\left\|\frac{\D\mathbb{P}_\theta}{\D\mathbb{P}}-\frac{\D\mathbb{P}_{\theta_*}}{\D\mathbb{P}}\right\|_{L^2(\mathcal{Y})}^2 \prec \left\|\frac{\D\mathbb{P}_\theta}{\D\mathbb{P}}-\frac{\D\mathbb{P}_{\theta_*}}{\D\mathbb{P}}\right\|_{\widetilde{\mathcal{Y}}}^2 + \mathbb{P}(y\in\widetilde{\mathcal{Y}})\prec \|\theta-\theta_*\|_\infty^2+ \mathbb{P}(y\in\widetilde{\mathcal{Y}})$$
Taken together, this implies A2. A3 can be verified analogously, starting with a Taylor expansion yielding $\mathbb{E}[l(\theta, \lambda, Y)-l(\theta, \lambda_*^\theta, Y)]=-\int f_*''(\widetilde{\lambda})(\lambda-\lambda_*^\theta)^2\D\mathbb{P} \asymp - \|\lambda-\lambda_*^\theta\|^2_{L^2(\mathcal{Y})}$ for some $\widetilde{\lambda}\in\Lambda$ on a path from $\lambda_*^\theta$ to $\lambda$.
\end{proof}
\subsection{Proof of Proposition \ref{thm:fdivnormality}}

\begin{proof}
First, we verify the conditions of Theorem \ref{thm:neuralAnormality}.  Note that $$l'(\theta, \lambda, Y_i)[v, w]=\int \lambda(y)\nabla_{\theta\to v}\frac{\D\mathbb{P}_\theta(y)}{\D\mathbb{P}(y)}\D\mathbb{P}(y) + \int w(y) \frac{\D\mathbb{P}_\theta(y)}{\D\mathbb{P}(y)}\D\mathbb{P}(y) - f_*'(\lambda(Y_i))w(Y_i)$$
The Lipschitz condition A4 is therefore satisfied by the Lipschitzness of $f_*'$, $\frac{\D\mathbb{P}_\theta}{\D\mathbb{P}}$ and that of $\nabla_{\theta\to v}\frac{\D\mathbb{P}_\theta}{\D\mathbb{P}}$. Towards A6 i), we apply a Taylor expansion with 2nd order mean-value reminder around $\lambda=\lambda_*^\theta$, yielding for some $\widetilde{\lambda}^\theta$ on a path between $\lambda_*^\theta$ and $\lambda$:
\begin{align*}
    & \mathbb{E}[l'(\theta, \lambda, Y_i)[v, \lambda_*^{\prime\theta}[v]]-l'(\theta, \lambda_*^\theta, Y_i)[v, \lambda_*^{\prime\theta}[v]]]\\
    &=\nabla_{\theta\to v}\nabla_{\lambda_*^\theta\to\lambda-\lambda_*^\theta}\mathbb{E}[l(\theta, \lambda_*^\theta, Y)] + \mathbb{E}[f_*''(\widetilde{\lambda}^\theta(Y))(\lambda-\lambda_*^\theta)^2w(Y)] \prec \mathbb{E}[(\lambda-\lambda_*^\theta)^2]
\end{align*}
where we used $\nabla_{\lambda_*^\theta\to\lambda-\lambda_*^\theta}\mathbb{E}[l(\theta, \lambda_*^\theta, Y)]\equiv 0$ to get rid of the first-order term. The last line follows from the boundedness (in absolute value) of $f_*''(\cdot)$ and $w(\cdot)$ on their compact support. Hence A6i) is satisfied. A7i) follows from:
$$\mathbb{V}[l'(\theta, \lambda, Y_i)[v, w]-l'(\theta, \lambda_*^\theta, Y_i)[v, w]]=\mathbb{V}[(f_*'(\lambda(Y))-f_*'(\lambda_*^\theta(Y)))w(Y)]\prec\mathbb{E}[(\lambda-\lambda_*^\theta)^2]$$
where the last step used the Lipschitzness of $f_*'$ and again the boundedness of $w(\cdot)$. Assumption A7 ii) is satisfied since $\mathcal{\Theta}_*$ is Donsker by assumption. Finally, we verify A6 ii). Applying the mean value theorem twice, we get that for some $\widetilde\theta, \widetilde\theta'$ on a path between $\theta_*$ and $\theta$, $\mathbb{E}[l'(\theta, Y)[v_*]-l'(\theta_*, Y)[v_*] -  \langle \theta-\theta_* , v_*\rangle = \nabla_{\widetilde\theta \to \widetilde\theta'-\theta_*}\nabla_{\widetilde\theta \to \theta-\theta_*} \mathbb{E}[l'(\widetilde\theta, Y)[v_*]]$ which is dominated by $\mathbb{E}[l(\theta, Y)]=D_f(\mathbb{P}_\theta \|\mathbb{P}_{\theta_*})$ via the last assumption stated in the proposition. Therefore A6ii) is satisfied, and Theorem 2.4 applies.
\end{proof}
\subsection{Proof of Proposition \ref{thm:cmrnormality}}
\begin{proof}
Note that $V=\nabla_{\theta_*}\nabla_{\theta_*'}\mathbb{E}[l(\theta_*, Y)]=\mathbb{V}[\nabla_{\theta_*}l(\theta_*, Y)]$. We verify the conditions of Theorem \ref{thm:neuralAnormality}, for $v_*=V^{-1}\zeta$, such that its conclusion becomes $\sqrt{n}\langle \theta-\theta_*, v_*\rangle=\sqrt{n}(\theta-\theta_*)'\zeta\to\mathcal{N}(0,\zeta'V^{-1}\zeta)$, which yields the Proposition via the Cramér-Wold device. Note that $l'(\theta,\lambda,Y)[v,w]=v'd(X,\theta)'\lambda(Z)+m(X,\theta)'w(Z)-\frac{1}{2}\lambda(Z)'w(Z)$ 
Hence assumption A4 follows from boundedness and Lipschitzness of $d(X,\cdot), m(X,\cdot)$. A5 holds by assumption. To verify A6i), notice that:
$$\mathbb{E}[l'(\theta,\lambda,Y)[v_*, \lambda_*^{\prime\theta}[v_*]]-l'(\theta,Y)]  = \mathbb{E}\left[\left(v_*'d(X,\theta)-\frac{1}{2}\lambda_*^{\prime\theta}[v_*](Z)\right)\left(\lambda-\lambda_*^\theta\right)(Z)\right]=0$$
where the last equality used the fact that $\mathbb{E}[v_*'d(X,\theta)|Z]=\frac{1}{2}\lambda_*^{\prime\theta}[v_*](Z)$.
Towards Assumption A6ii), note that we can apply a first-order Taylor expansion with mean-value reminder twice, which yields for some $\Bar{\theta}, \widetilde{\theta}$ on a path between $\theta_*$ and $\theta$: $$\mathbb{E}[l'(\theta, Y)[v_*]-l'(\theta_*, Y)[v_*]-\langle \theta-\theta_*, v_*\rangle]=(\Bar{\theta}-\theta_*)'\nabla_{\widetilde{\theta}}\nabla_{\widetilde{\theta'}}\mathbb{E}[l'(\widetilde{\theta}, Y)][v_*](\theta-\theta_*)\prec \|\theta-\theta_*\|_2^2$$
Given the identification assumption \ref{eq:cmr}, we can use a second-order Taylor expansion with mean-value reminder to show that $\|\theta-\theta_*\|_2^2 \asymp \mathbb{E}[l(\theta, Y)-l(\theta_*, Y)]$, which then yields A6ii). Similarly, we can show A7ii) via a mean-value reminder: 
$$\mathbb{E}[(l'(\theta, Y)[v]-l'(\theta_*, Y)[v])^2]=\mathbb{E}\left[\left((\theta-\theta_*)'\nabla_{\widetilde{\theta}}l'(\widetilde{\theta}, Y)[v]\right)^2\right]\prec\|\theta-\theta_*\|_2^2$$
We similarly can establish A7i) via 
\begin{equation*}
\mathbb{V}[l'(\theta, \lambda, Y)[v,\lambda_*^{\prime\theta}[v_*]]-l'(\theta, \lambda_*^\theta, Y)[v,\lambda_*^{\prime\theta}[v_*]]]\prec \mathbb{E}[(\lambda(Z)-\lambda_*^\theta(Z))^2]\asymp\mathbb{E}[l(\theta,\lambda,Y)-l(\theta,\lambda_*^\theta, Y)]
\end{equation*}
\end{proof}
\subsection{Proof of Proposition \ref{thm:sbeedrates}}
\begin{proof}
A0 holds by assumption, and condition A1 is implied by Lipschitzness of $V_\theta, P_\theta$ in $\theta$. Continuity of $V_\theta, P_\theta$, compactness of $\Theta$ and A0 further imply that there is some $0\leq M<\infty$ such that $$\sup_{s,a,s_+,\theta}|R_\theta(s,a)+\beta V(s^+)-V(s)-\log P_\theta(a|s)|\leq M$$ Given that $\lambda_*^{\theta_*}\equiv 0\implies l(\theta_*, Y)=0$, condition A2 then follows from $$\mathbb{V}[l(\theta, Y)-l(\theta_*, Y)]\prec \mathbb{E}[l(\theta, Y)^2]\leq 2 M^2\mathbb{E}[\lambda_*^\theta(s,a)^2]\prec \mathbb{E}[l(\theta, Y)]=\mathbb{E}[l(\theta, Y)-l(\theta_*, Y)]$$
as well as $\mathbb{E}[l(\theta, Y)-l(\theta_*, Y)]\prec\mathbb{E}[(\lambda_*^\theta(s,a)-\lambda_*^{\theta_*}(s,a))^2|(s,a)\in\widetilde{\mathcal{X}}]+\mathbb{P}((s,a)\not\in\widetilde{\mathcal{X}})\prec\|\theta-\theta_*\|^2_{\widetilde{\mathcal{X}}}+\mathbb{P}((s,a)\not\in\widetilde{\mathcal{X}}), \ \forall\widetilde{\mathcal{X}}\subset\Bar{\mathcal{X}}$. A3 can be established analogously, hence the conclusions of Theorem \ref{thm:neuralArate} hold.
\end{proof}
\subsection{Proof of Proposition \ref{thm:rieszrate}}
\begin{proof}
Note that $\lambda_*^\theta(x)=\theta(x)-\theta_*(x)$ follows from the first order conditions of the adversary. Condition A0 is satisfied by assumption, and A1 follows from Lipschitzness of $m(Y,\cdot)$ and boundedness. Lipschitzness of $m(\theta,\lambda(x))$ in $\lambda(x)$ and boundedness imply that $l(\theta,\lambda,Y)=m(Y,\lambda)-\theta(x)\lambda(x)-\lambda(x)^2/2$ is Lipschitz in $\lambda(x)$. This implies
$$\mathbb{V}[l(\theta,\lambda,Y)-l(\theta,\lambda_*^\theta,Y)]\prec \mathbb{E}[(\lambda(x)-\lambda_*^\theta(x))^2]$$
and together with $\lambda_*^\theta(x)=\theta(x)-\theta_*(x)$, it yields:
$$\mathbb{V}[l(\theta_*,\lambda_*^{\theta_*},Y)-l(\theta,\lambda_*^\theta,Y)]\prec \mathbb{E}[(\theta(x)-\theta_*(x))^2]$$
Both bounds imply A3 and A2 respectively. The result follows because the loss $\mathbb{E}[l(\theta,Y)]$ is proportional to the squared L2 norm of $\theta-\theta_*$.
\end{proof}
\end{document}